\documentclass[preprint,10pt]{elsarticle}
\pdfoutput=1
\usepackage{graphicx}
\usepackage{caption}
\usepackage[labelformat=simple]{subcaption}

\usepackage{amsmath}
\usepackage{amssymb}
\usepackage{amsthm}
\usepackage{setspace}
 \usepackage[normalem]{ulem} 
\usepackage{cancel}
\usepackage[linesnumbered,ruled,vlined,noend]{algorithm2e}
\usepackage{algpseudocode}
\usepackage{color}
\usepackage[]{color}
\usepackage[]{url}
\usepackage{hyperref}
\usepackage{cleveref}
\usepackage{enumerate}
\usepackage{enumitem}
\usepackage{multirow}
\usepackage{makecell} 

\usepackage{tikz}
\usepackage{standalone}
\usetikzlibrary{arrows}
\usetikzlibrary{decorations} 
\usetikzlibrary{decorations.markings}
\usetikzlibrary{decorations,fit,shapes} 
\usetikzlibrary{arrows.meta,positioning}
\usetikzlibrary{shapes.misc}

\newcommand{\prefa}{\mbox{{\sf Pref($a$)}}}
\newcommand{\preflqa}{\mbox{{\sf PrefLQ($a$)}}}
\newcommand{\prefb}{\mbox{{\sf Pref($b$)}}}
\newcommand{\prefh}{\mbox{{\sf Pref($h$)}}}

\newcommand{\prefsa}{\mbox{{\sf PrefS($a$)}}}

\newcommand{\prefslqa}{\mbox{{\sf PrefSLQ($a$)}}}
\newcommand{\prefu}{\mbox{{\sf Pref($u$)}}}

\newcommand{\HRLQ}{\mbox{{\sf HRLQ}}}

\newcommand{\RSM}{\mbox{{\sf RSM}}}
\newcommand{\CRSM}{\mbox{{\sf CRITICAL-RSM}}}

\newcommand{\Df}[1]{\mbox{{\sf $def(#1)$}}}
\newcommand{\Dfa}[1]{\mbox{{\sf $def_{\mathcal{A}}(#1)$}}}
\newcommand{\Dfb}[1]{\mbox{{\sf $def_{\mathcal{B}}(#1)$}}}

\newcommand{\HRTMSLQ}{{\sf{HRT}\text{-}\sf{MSLQ}}}
\newcommand{\HH}{\mathcal{H}}
\newcommand{\RR}{\mathcal{R}}
\renewcommand{\AA}{\mathcal{A}}
\newcommand{\BB}{\mathcal{B}}

\renewcommand{\S}{{s}}
\newcommand{\T}{{t}}

\newtheorem{property}{Property}
\newtheorem{cl}{Claim}
\newtheorem{definition}{Definition}
\newtheorem{theorem}{Theorem}
\newtheorem{lemma}[theorem]{Lemma}
\newtheorem{observation}{Observation}
\newtheorem{remark}{Remark}

\definecolor{lightgray}{rgb}{0.83, 0.83, 0.82}

\tikzset{cross/.style={cross out, draw=black, minimum size=2*(#1-\pgflinewidth), inner sep=0pt, outer sep=0pt},
	cross/.default={1pt}}
\usetikzlibrary{arrows, decorations.pathmorphing}

\tikzstyle{vertex}=[auto=left,circle,draw=black!80,fill=none,minimum size=15pt,inner sep=0pt]

\tikzset{
    photon/.style={decorate, decoration={snake}, draw=red}}

\newcommand{\etal}{\textit{et al}.}

\newenvironment{appendix-lemma}[1]{\vspace{0.1in}\noindent{\bf Lemma~#1~} \em }{\vspace{0.1in}}
\newenvironment{appendix-claim}[1]{\vspace{0.1in}\noindent{\bf Claim~#1~} \em }{\vspace{0.1in}}
\newenvironment{appendix-theo}[1]{\vspace{0.1in}\noindent{\bf Theorem~#1~} \em }{\vspace{0.1in}}
\newenvironment{appendix-prop}[1]{\vspace{0.1in}\noindent{\bf Property~#1~} \em }{\vspace{0.1in}}

\newenvironment{proced}[1][htb]{%
    \begin{algorithm}[#1]%
}{%
    \end{algorithm}%
}

\algrenewcommand{\algorithmiccomment}[1]{\hfill\hfill\hfill\hfill\hfill\hfill\hfill// #1}

\renewenvironment{proof}{{\textit{Proof:}}}{\qed}
\newcommand{\proofofref}{}
\newproof{zproofof}{Proof of \proofofref}

\usepackage[margin = 1.2in]{geometry}

\title{Critical Relaxed-Stable Matchings with Ties in the Many-to-Many Setting\tnoteref{label1}}
\tnotetext[]{A preliminary version~\cite{nasre2023critical} of this work appeared in 49th International Workshop on Graph-Theoretic Concepts in Computer Science 2023 (WG 2023)} 

\journal{Discrete Applied Mathematics}
\begin{document}

\begin{frontmatter}



\author[1]{Meghana Nasre}
\ead{meghana@cse.iitm.ac.in}
\affiliation[1]{organization={Department of CSE},
            addressline={IIT Madras},
            city={Chennai},
            postcode={600036}, 
            state={Tamil Nadu},
            country={India}}
            
\author[2]{Prajakta Nimbhorkar}
\ead{prajakta.nimbhorkar@gmail.com}

\affiliation[2]{organization={Computer Science},
            addressline={Chennai Mathematical Institute and UMI ReLaX},
            city={Chennai},
             postcode={603103}, 
            state={Tamil Nadu},
            country={India}}

\author[3]{Keshav Ranjan\fnref{fn3}}  
\ead{ranjankeshav08@gmail.com}
\affiliation[3]{organization={Computer Science},
            addressline={Chennai Mathematical Institute},
            city={Chennai},
            postcode={603103}, 
            state={Tamil Nadu},
            country={India}}
     \fntext[fn3]{This work was done while the author was at IIT Madras.}
 


\begin{abstract}
We study the many-to-many bipartite matching problem in the presence of preferences where ties, as well as lower quotas, may appear on both sides of the bipartition. Each vertex has a positive upper quota limiting the number of vertices it can be assigned to, a non-negative lower quota specifying its desired minimum, and a preference ordering, possibly containing ties, over its neighbors. A \emph{critical} matching is a matching that fulfills vertex lower quotas to the maximum possible extent in aggregate. We seek to compute a matching that is both critical and optimal with respect to the preferences of the vertices.

Stability, a well-accepted notion of optimality in the presence of two-sided preferences, is generalized to weak stability in the presence of ties. However, a matching that is both critical and weakly stable need not exist. We show that deciding whether such a matching exists is NP-complete, even in the one-to-one setting with complete preference lists and lower quotas on only one side of the bipartition. Popularity is another well-investigated notion of optimality for the two-sided preference model; however, a popular matching need not exist in the presence of ties, even without lower quotas. We therefore generalize the notion of \emph{relaxed stability}, introduced by Krishnaa, Limaye, Nasre, and Nimbhorkar (JoCO 2023), to our setting. We prove that a critical relaxed-stable matching always exists, although computing one of maximum size is NP-hard. 
Our polynomial-time algorithm computes a critical relaxed-stable matching whose size is at least $\frac{2}{3}$ times that of a maximum-size relaxed-stable matching, even when the latter is not critical.
\end{abstract}


\end{frontmatter}

\section{Introduction} \label{sec:intro}

Preference-based many-to-many matching problems model applications such as assigning students to courses~\cite{CEFMMP2014} and workers to firms~\cite{roth1984stability}, where agents on both sides have preferences over potential partners and may accept multiple partners. In these applications, preferences must be considered alongside minimum participation requirements. For example, students typically need to complete a minimum number of courses each semester, while courses may have minimum enrollment requirements. Similarly, in the worker-firm assignment, certain workers must be assigned jobs, and firms may need a minimum number of workers for their operations. These minimum requirements can be represented by \emph{lower quotas}, specifying the desired minimum number of partners for each agent. Therefore, it is natural to have lower quotas on both sides of the bipartition. Lower quotas~\cite{BFIM10,HIM16,NN17} have been considered in the literature in the context of matching residents to hospitals (\textit{e.g.} the National Resident Matching Program in the U.S.A.), motivated in part by the understaffing of rural hospitals~\cite{roth84,Roth86}. Another practical consideration is the presence of \emph{ties} in preferences. Hospitals with many applicants often find it difficult to specify strict preference lists~\cite{swat/IrvingMS00}, and course instructors may be indifferent among students with equal scores. These applications motivate studying lower quotas and ties together in the many-to-many setting.

In this work, we study many-to-many matching with ties in preferences and lower quotas on both sides of the bipartition. Our setting generalizes several models, such as the one-to-one setting of the stable marriage problem~\cite{GS62,irving1994stable}, the many-to-one setting of the Hospital/Residents problem with ties~\cite{swat/IrvingMS00}, and the many-to-many assignment problem~\cite{brandl2019two} -- all of these works are without lower quotas. Previous work has studied lower quotas with strict
preferences~\cite{BFIM10,HIM16,NN17,Yokoi20,NNRS21,Kavitha2021,NNRS2024popular}, as well as lower quotas with ties in the many-to-one setting, where positive lower quotas are restricted to hospitals~\cite{GokoMMY22,MakinoMY22}. The conference version of our work~\cite{nasre2023critical} considers ties and lower quotas on both sides in the one-to-one setting. To the best of our knowledge, our conference paper was the first to study ties and lower quotas on both sides of the bipartition. Independently and in parallel, Cs{\'a}ji~\cite{csaji2024AAMAS} studied a more general one-to-one framework incorporating critical vertices, critical edges, and free edges. Subsequently, Cs{\'a}ji \etal~\cite{Simple15Approximation} extended this framework to incorporate matroid constraints on the vertices. Here, we develop a direct proposal-based algorithm for the many-to-many setting with ties and lower quotas on both sides. It builds on the deferred-acceptance approach of Gale and Shapley~\cite{GS62} and uses explicit proposal and acceptance rules, making it straightforward to implement.

Formally, the input is a bipartite graph $G = (\AA \cup \BB, E)$, where $\AA$ and $\BB$ are disjoint sets of vertices, and $E$ consists of acceptable vertex pairs. Each vertex $u \in \AA\cup\BB$ specifies a \emph{preference list}, denoted by $\prefu$,  ranking its neighbors in $G$. Smaller ranks indicate more-preferred neighbors, and neighbors assigned the same rank form a \emph{tie}. A preference list is \emph{strict} if it contains no ties. For two neighbors $v_1$ and $v_2$ of $u$, we write $v_1 \succ_u v_2$ if $u$ strictly prefers $v_1$ to $v_2$, and $v_1 \succeq_u v_2$ if $u$ either strictly prefers $v_1$ to $v_2$ or is indifferent between them. Each vertex $u$ also has a lower quota $q^-(u) \in \mathbb{Z}^+ \cup\{0\}$ and an upper quota $q^+(u)\in\mathbb{Z}^+$, with $q^-(u)\le q^+(u)$. The upper quota specifies the maximum number of partners that may be assigned to $u$, while the lower quota specifies its desired minimum. A vertex $u$ is an \emph{lq}-vertex if $q^-(u)>0$ and a non-\emph{lq} vertex otherwise.

Following the terminology in~\cite{brandl2019two,HIM16}, a \emph{matching} $M$ in $G$ is a subset of $E$ such that $|M(u)| \le q^+(u)$ for every $u \in \AA\cup\BB$, where $M(u)$ denotes the set of neighbors assigned to $u$ in $M$. For a matching $M$, a vertex $u \in \AA\cup\BB$ is \emph{fully-subscribed} if $|M(u)| = q^+(u)$, \emph{under-subscribed} if $|M(u)| < q^+(u)$, \emph{deficient} if $|M(u)| < q^-(u)$, and \emph{surplus} if $|M(u)| > q^-(u)$. A matching is \emph{feasible} if it satisfies every vertex's lower quota; that is, no vertex is deficient. However, a feasible matching need not exist~\cite{NNRS2024popular}. We therefore seek a \emph{critical matching}, which fulfills lower quotas to the maximum possible extent in aggregate. Formally, the deficiency of a vertex $u$ in a matching $M$ is $\max\{0, q^-(u)-|M(u)|\}$, and the total deficiency of $M$, denoted by $\Df{M}$, is the sum of the deficiencies of all vertices in $\AA \cup \BB$. That is,
\[
\Df{M} = \sum_{u\in\AA\cup\BB}\max\{0,q^-(u)-|M(u)|\}.
\]

\begin{definition}[Critical Matching]\label{def:critical}
A matching $M$ in $G$ is critical if there is no matching $N$ in $G$ with $\Df{N} < \Df{M}$. 
\end{definition}

In this work, we are interested in a critical matching that is also optimal with respect to the vertices' preferences. Two extensively studied notions of optimality for bipartite matching with two-sided preferences are \emph{stability} and \emph{popularity}.

\begin{definition}[Stable Matchings]\label{def:bPT}
Given a matching $M$, a pair $(a, b) \in E \setminus M$ is a blocking pair with respect to $M$ if $(i)$ either $|M(a)| < q^+(a)$ or $b\succ_a b'$ for some $b' \in M(a)$, and $(ii)$ either $|M(b)| < q^+(b)$ or $a\succ_b a'$ for some $a'\in M(b)$. The matching $M$ is stable if there are no blocking pairs for it.
\end{definition}

Thus, each endpoint of a blocking pair is either
under-subscribed or {\em strictly} prefers the other endpoint to one of its current partners. In the presence of ties, this notion is called \emph{weak stability}~\cite{irving1994stable}. Two other notions, \emph{strong stability} and \emph{super stability}, have also been studied, but matchings satisfying them need not exist, even without lower quotas. Throughout this paper, we use \emph{stability} to mean weak stability.

\medskip

\noindent\textbf{Stable matching in the presence of ties and \emph{lq}-vertices:}
Every stable marriage instance with ties admits a stable matching, which can be computed in polynomial time~\cite{irving1994stable}. With strict preferences, all stable matchings have the same size. With ties and incomplete preference lists, their sizes may differ, and computing a maximum-size or minimum-size stable matching is NP-hard~\cite{manlove2002hard}. Lower quotas introduce a further difficulty. A matching that is both stable and critical need not exist, even with strict preferences~\cite{NNRS2024popular}. In the presence of ties, we show that deciding whether such a matching exists is NP-complete, even in the one-to-one setting with complete preference lists and positive lower quotas only on one side of the bipartition (Theorem~\ref{cl:nphardCS}, Section~\ref{subsec:NPhard}).

\medskip

\noindent\textbf{Popular matching in the presence of
\emph{lq}-vertices:}
Popularity provides another way to account for preferences. Informally, a matching is \emph{popular} within a set of matchings if it never loses a pairwise election against another matching in that set, where vertices cast votes. For strict preferences and lower quotas, this notion has been extensively studied~\cite{NN17,NNRS21,Kavitha2021,NNRS2024popular}; and it is known that a matching that is popular within the set of critical matchings always exists and can be computed in polynomial time~\cite{Kavitha2021,NNRS21,NNRS2024popular}. However, with ties, a popular matching need not exist, even without lower quotas, and deciding existence is NP-hard~\cite{biro2010popular}. This hardness holds even when ties occur only on one side of the bipartition~\cite{cseh2017popular}. These limitations motivate considering \emph{relaxed stability}.

\medskip

\noindent {\bf Relaxed stability in the presence of ties and \emph{lq}-vertices: } Krishnaa \etal~\cite{krishnaa2023envy} introduced and studied relaxed stability for the Hospital/Residents problem with lower quotas (\HRLQ). In this setting, preferences are strict, each resident has upper quota one and lower quota zero, and each hospital $h$ has an upper quota $q^+(h)$ and a lower quota $q^-(h)\le q^+(h)$. To minimize deficiency, some residents may need to be assigned to \emph{lq}-hospitals even when they prefer other hospitals. Relaxed stability permits a resident to participate in a blocking pair only if its assigned hospital is non-surplus. Thus, reassigning such a resident to another hospital would increase the deficiency of its current hospital.

\begin{definition}[Relaxed Stability in \HRLQ~\cite{krishnaa2023envy}]\label{def:rsmHRLQ}
A matching $M$ is relaxed-stable if no unmatched resident participates in a blocking pair and, for every resident $r$ who participates in a blocking pair, its assigned hospital $h$ satisfies $|M(h)|\le q^-(h)$.
\end{definition}

A matching is a \emph{critical relaxed-stable matching} (\CRSM) if it is both critical and relaxed-stable. 

\medskip

\noindent\textbf{Our contribution:} We generalize relaxed stability to the many-to-many setting with ties in preferences and lower quotas on both sides of the bipartition, and prove that a critical relaxed-stable matching always exists. Our main result is a polynomial-time algorithm that computes such a matching whose size is at least $\frac{2}{3}$ times the size of a maximum-size relaxed-stable matching. Thus, our algorithm minimizes total deficiency while its size guarantee is measured against \emph{all} relaxed-stable matchings, including those that are not critical.

\begin{theorem}\label{theo:main}
For every many-to-many instance $G=(\AA\cup\BB,E)$ with ties in preferences and lower quotas on both sides of the bipartition, there is a polynomial-time algorithm that computes a critical relaxed-stable matching $M$ satisfying $|M| \ge \frac{2}{3}|M_{max}|$, where $M_{max}$ is a maximum-size relaxed-stable matching in $G$.
\end{theorem}

Computing a maximum-size \CRSM\ is NP-hard even when every vertex has upper quota one and lower quota zero. In this case, every matching is critical, and relaxed stability coincides with weak stability~\cite{manlove2002hard}. Our approximation guarantee matches the $\frac{2}{3}$ bound for maximum-size stable matching with ties~\cite{mcdermid20093,Kiraly13,paluch2014faster}. Moreover, Dudycz \etal~\cite{dudycz2022tight} show that, under the Small Set Expansion Hypothesis~\citep{RaghavendraSmallSet}, achieving a $(\frac{2}{3}+\epsilon)$-approximation is NP-hard for every fixed $\epsilon>0$. Consequently, assuming this hypothesis and $P \ne NP$, the approximation factor $\frac{2}{3}$ is optimal.

We further prove that our output satisfies the many-to-many extension of the stronger notion of critical relaxed stability introduced by Cs{\'a}ji~\cite{csaji2024AAMAS} (Remark~\ref{remark:Csaji} and Lemma~\ref{lem:equivCsaji}). Thus, the same algorithm guarantees this stronger property together with criticality and the size bound in Theorem~\ref{theo:main}.

We also establish that deciding whether a stable critical matching exists is NP-complete, even in the one-to-one setting with complete preference lists and positive lower quotas only on the hospital side. Hardness holds even when ties occur only on one side, with at most one tie of length two per list (Theorem~\ref{cl:nphardCS}). Our proof adapts the auxiliary-hospital construction of Goko \etal~\cite{GokoMMY22}.

Our algorithm combines Kir{\'a}ly's framework for handling ties~\cite{Kiraly13} with the multi-level proposal framework of Nasre \etal~\cite{NNRS2024popular} for handling lower quotas. A key idea in Kir{\'a}ly's algorithm is the notion of an \emph{uncertain proposal}, which plays a crucial role in handling ties on the proposer side and obtaining the $\frac{2}{3}$-approximation guarantee.

As a first technical step, we refine this notion and adapt Kir{\'a}ly's framework to the many-to-many setting. In doing so, we address a technical issue in the definition of uncertain proposals in the many-to-one algorithm~\cite{Kiraly13}, which, among other conditions, requires the proposing vertex to be fully subscribed. An example in~\ref{append:kiralyM1} shows that imposing this condition can lead to an unstable matching. Our refinement removes this requirement while retaining the underlying approach. A $\frac{2}{3}$-approximation for the many-to-many setting without lower quotas is already known~\cite{paluch2014faster}; our formulation enables us to integrate Kir{\'a}ly's approach with the multi-level framework for lower quotas.

\medskip

\medskip
\noindent {\bf Challenges in our setting:}
In case of strict preferences and lower quotas, Nasre \etal~\cite{NNRS2024popular} use a multi-level proposal framework with dynamically changing levels and capacities, together with level-dependent preferences. In case of ties without lower quotas and many-to-one setting, Kir{\'a}ly \cite{Kiraly13} introduced the notion of uncertain proposals. We combine both settings and additionally, generalize it to the many-to-many setting. As a result, the following challenges arise:
\begin{enumerate}
    \item In Kir{\'a}ly's hospital-proposing algorithm~\cite{Kiraly13}, a hospital may be uncertain about its proposals to several residents at once, but each resident holds at most one proposal. In our setting, both proposer and receiver may simultaneously have several uncertain proposals. Thus the receiver needs to make a careful choice of which proposal to reject.
    \item Kir{\'a}ly's algorithm~\cite{Kiraly13} deletes a neighbor $b$ from $\prefa$ when $b$ rejects $a$, unless the proposal is uncertain. We cannot do this due to the multi-level proposal framework required for lower quotas. For this, we require a significantly technical definition of {\em favorite neighbor} and the additional  book-keeping of {\em marked neighbor}. 
    \item We need to prove that the output matching is both critical and relaxed-stable, and establish the $\frac{2}{3}$ size guarantee. For this, we define a {\em cloned graph}, which is not needed in \cite{Kiraly13}.
\end{enumerate}
\noindent{\bf Related work:}
For strict preferences and lower quotas, various relaxations of stability, including envy-freeness~\cite{Yokoi20,krishnaa2023envy}, popularity~\cite{NN17,Kavitha2021,NNRS21,NNRS2024popular}, and relaxed stability~\cite{krishnaa2023envy}, have been studied. Envy-freeness is defined by the absence of envy-pairs. However, a critical envy-free matching need not exist~\cite{krishnaa2023envy}. Relaxed stability and popularity are incomparable, even in the one-to-one setting with strict preferences and \emph{lq}-vertices restricted to one side of the bipartition, as pointed out by Krishnaa \etal~\cite{krishnaa2023envy}. Hamada~\etal~\cite{HIM16} consider computing a matching with the minimum number of blocking pairs or blocking residents among all \textit{feasible} matchings for the \HRLQ\ problem.

For the maximum-size stable marriage problem with ties and incomplete preference lists, without \emph{lq}-vertices, a long line of work investigates approximation algorithms for the general problem and various restricted settings~\cite{kiraly2011better,mcdermid20093,Kiraly13,paluch2014faster,iwama201425,huang2015improved}. The best-known approximation guarantee for one-sided ties is due to Lam and Plaxton~\cite{lam20191}. For two-sided ties, the best-known guarantee is $\frac{2}{3}$, attained by algorithms of McDermid, Kir{\'a}ly, and Paluch~\cite{mcdermid20093,Kiraly13,paluch2014faster}. Dudycz \etal~\cite{dudycz2022tight} established that, assuming the Small Set Expansion Hypothesis~\citep{RaghavendraSmallSet}, approximating the maximum-size weakly stable matching problem with ties and incomplete preference lists within a factor of $\frac{2}{3} + \epsilon$ is NP-hard for every fixed $\epsilon > 0$.

Our model allows ties in preferences and \emph{lq}-vertices on both sides of the bipartition in the many-to-many setting. Related works by Goko \etal~\cite{GokoMMY22} and Makino \etal~\cite{MakinoMY22} consider the many-to-one setting, with positive lower quotas restricted to hospitals. Goko \etal\ assume complete preference lists, whereas Makino \etal\ allow incomplete lists. They study the \textit{Hospitals/Residents problem with Ties to Maximally Satisfy Lower Quotas} (\HRTMSLQ), which seeks a weakly stable matching maximizing the total \emph{satisfaction ratio} of hospital lower quotas. For a hospital $h$ with $q^-(h) > 0$, its satisfaction ratio is $\min\{1, \frac{|M(h)|}{q^-(h)}\}$; it is defined to be $1$ when $q^-(h) = 0$. Since a weakly stable matching always exists, \HRTMSLQ\ always admits a solution, whereas a stable critical matching need not exist. These works require stability and optimize lower-quota satisfaction subject to this requirement. Consequently, their output is guaranteed to be stable but need not be critical. In our work, we minimize total deficiency over all matchings and guarantee that the output is both critical and relaxed-stable.
\section{Preliminaries} \label{sec:prelim}

Our proposal-based algorithm combines ideas from (i) Kir{\'a}ly's algorithm~\cite{Kiraly13} for computing a $\frac{2}{3}$-approximation to a maximum-size stable matching with two-sided ties and no lower quotas, and (ii) the multi-level algorithm for computing a popular critical matching with strict preferences and lower quotas on both sides of the bipartition~\cite{NNRS2024popular}. We first describe an adaptation of Kir{\'a}ly's algorithm to the many-to-many setting.

\subsection{Kir{\'a}ly's algorithm for the many-to-many setting}\label{sec:GenKiraly}

In this section, we extend the many-to-one variant of Kir{\'a}ly's algorithm~\cite{Kiraly13} to the many-to-many setting to compute a $\frac{2}{3}$-approximation to a maximum-size stable matching. Vertices in $\AA$ propose, and vertices in $\BB$ accept or reject proposals. As long as a vertex $a \in \AA$ is under-subscribed and has an unproposed or marked neighbor, it proposes to its \emph{favorite} neighbor (Definition~\ref{def:favNbr}); marking is explained below. If $a$ remains under-subscribed after processing all unproposed and marked neighbors, it attains \emph{starred status}, denoted by $a^*$, and restarts proposing from the beginning of $\prefa$.

Starred status gives $a$ priority over unstarred vertices tied with it in each neighbor's preference list. In other words, starred status can be interpreted as improving (equivalently decreasing) the rank of $a$ in every neighbor's preference list by a fixed amount $0 < \epsilon < 1$. Thus, a neighbor $b$ prefers $a^*$ to any unstarred vertex tied with $a$ in $\prefb$, while every vertex strictly preferred to $a$ in $\prefb$ continues to be preferred over $a^*$.

To describe the acceptance and rejection rules, we first introduce \emph{uncertain proposals}.

\begin{definition}[Uncertain Proposal] \label{def:uncertainProp}
Let $M$ be the current matching when an unstarred vertex $a\in\AA$ proposes to a rank-$k$ neighbor $b\in\BB$ in $\prefa$. If accepted, this proposal is labeled \emph{uncertain} if it is the first proposal from $a$ to $b$ and there exists another rank-$k$ neighbor $b'\ne b$ that is unproposed by $a$ and satisfies $|M(b')|<q^+(b')$.
\end{definition}

In the many-to-many setting, a vertex may have several accepted proposals labeled uncertain, and thus, such proposals must be handled carefully. A vertex $a$, before moving to a less-preferred rank, say $k+1$, labels all its remaining uncertain proposals at the current rank $k$ as `not uncertain'. It also clears any remaining uncertainty labels before attaining starred status.

\medskip

\noindent\textbf{Accept/reject rules:} Each time $a$ or $a^*$ proposes to its favorite neighbor $b$, the following rules apply. If $a^*$ proposes to $b$ while $a$ is already matched to $b$, we replace $a$ by $a^*$ in $M(b)$ without changing the underlying matching. Otherwise:

\begin{enumerate}[label=R\arabic*., ref=R\arabic*]
    \item\label{itm:AccRej1} If $b$ is under-subscribed, it accepts the proposal.

    \item\label{itm:AccRej2} If $b$ is fully subscribed and holds an uncertain proposal from some $a'\in M(b)$, it rejects $a'$ and accepts the incoming proposal, regardless of the relative ranks of $a$ and $a'$ in $\prefb$.

    \item \label{itm:AccRej3} If $b$ is fully subscribed and holds no uncertain proposal, it compares the proposer with a least-preferred partner in $M(b)$. To select such a partner, first consider the partners at the largest rank in $\prefb$. Among these, choose an unstarred partner if one exists, and a starred partner otherwise. Let $a'$ be the selected partner. If $b$ strictly prefers the proposer to $a'$, accounting for starred status, it rejects $a'$ and accepts the proposal; otherwise, it rejects the proposer.
\end{enumerate}

Kir{\'a}ly's algorithm~\cite{Kiraly13} deletes a neighbor $b$ from $\prefa$ when $b$ rejects $a$, unless the proposal is uncertain. In our case, we integrate this adaptation into a multi-level proposal framework, where vertices may change levels during execution. We therefore retain preference lists and track proposals without deleting neighbors. To identify neighbors that may be revisited, we introduce the notion of \emph{marking}, described below. This notion is also used to define the \emph{favorite neighbor} (Definition~\ref{def:favNbr}).

When $b$ rejects $a'$ under rule~(\ref{itm:AccRej2}) because of uncertain proposal, $a'$ marks $b$. If $a'$ remains under-subscribed after proposing to all unproposed neighbors at the rank of $b$, it must repropose to its marked neighbors at that rank before moving to a less-preferred rank. This is crucial for ensuring stability. A marked neighbor is unmarked when the proposal is made.  In contrast, rejections under rule~(\ref{itm:AccRej3}) do not create marks.

We now define the favorite neighbor, which uniquely specifies the next neighbor to receive a proposal, even when several neighbors share the same rank.

\begin{definition}[Favorite Neighbor of $a$] \label{def:favNbr}
Let $k$ be the best rank in $\prefa$ containing a neighbor that is unproposed or marked by $a$. The favorite neighbor $b$ of $a$ is selected as follows:
\begin{enumerate}[label=(\roman*)]
    \item \label{itm:FavNbrKiral1} If there is an under-subscribed neighbor at rank $k$ that is unproposed by $a$, choose the one with the smallest index.

    \item \label{itm:FavNbrKiral2} Otherwise, if there is a fully subscribed neighbor at rank $k$ that is unproposed by $a$, choose the one with the smallest index.

    \item \label{itm:FavNbrKiral3} Otherwise, choose the smallest-index neighbor marked by $a$ at rank $k$.
\end{enumerate}
The same definition applies to $a^*$, using its separate proposal history.
\end{definition}

Since $a^*$ restarts proposing from the beginning of $\prefa$, its favorite neighbor may already be matched to $a$. In this case, we replace $a$ with $a^*$ in $M(b)$ without changing the underlying pair $(a,b)$. Thus, $a$ and $a^*$ never occupy two positions of $b$ simultaneously. For example, consider an instance with a single edge $(a,b)$ and upper quotas $q^+(a) = q^+(b) = 2$. The first proposal from $a$ to $b$ is accepted. Since $a$ remains under-subscribed after exhausting its list, it attains starred status and proposes to $b$ again. This proposal by $a^*$ updates the starred status of the existing pair, and the algorithm terminates with $a^*$ matched to $b$, corresponding to the matching $M = \{(a, b)\}$.

This completes the description of our many-to-many adaptation of Kir{\'a}ly's algorithm, which computes a $\frac{2}{3}$-approximation to a maximum-size stable matching. Such an approximation guarantee for the many-to-many setting is already known~\cite{paluch2014faster}. We present this formulation to facilitate its integration into our multi-level algorithm for lower quotas. For completeness, we provide its full pseudocode and analysis in~\ref{append:kiralyMM}.

\subsection{Overview of the multi-level popular critical matching algorithm}
We briefly describe the algorithm of Nasre \etal~\cite{NNRS2024popular} for computing a maximum-size popular critical matching in the many-to-many setting with strict preferences and lower quotas on both sides of the bipartition. Let $\S$ and $\T$ denote the sums of lower quotas of vertices in $\AA$ and $\BB$, respectively, that is, $\S = \sum_{a \in \AA}q^-(a)$ and $\T = \sum_{b \in \BB}q^-(b)$. The algorithm uses $\S + \T +2$ levels, indexed from $0$ to $\S + \T + 1$. All vertices in $\AA$ start at level $0$ and may advance through these levels during execution. A vertex $a$ at level $\ell$ is denoted by $a^\ell$. A vertex $b \in \BB$ prefers $a_i^\ell$ to $a_j^{\ell'}$ if either (i) $\ell > \ell'$, or (ii) $\ell = \ell'$ and $a_i \succ_b a_j$.

At levels $0, \ldots, \T-1$, proposals prioritize the lower quotas of vertices in $\BB$. Each under-subscribed vertex $a \in \AA$ proposes only to \emph{lq}-vertices in its preference list and advances to the next level if it remains under-subscribed after exhausting this list at the current level. At level $\T$, it may propose to all its neighbors. If it remains under-subscribed after exhausting $\prefa$ at level $\T$, it advances to level $\T+1$ and proposes again until it becomes fully subscribed or exhausts its list. If it remains deficient after exhausting level $\T+1$, it advances to level $\T+2$. At these higher levels, it continues proposing while deficient, advancing after exhausting each list, until it becomes non-deficient or exhausts its list at level $\S + \T +1$. Whenever a vertex advances to a new level, it restarts from the beginning of the applicable preference list. These transitions occur separately for each vertex; vertices need not be processed at the same level. 

The capacities used for proposals depend on the proposer's level and the receiver's current partners, as summarized in Table~\ref{tab:quota}. Nasre \etal~\cite{NNRS2024popular} show that the resulting matching has maximum size among matchings that are popular within the set of critical matchings. 

\begin{table}
\centering
\renewcommand{\arraystretch}{1.3} 
\begin{tabular}{| p{2.9 cm} |p{1cm} | p{1.9 cm} |p{1cm} p{4cm}|}
\hline 
      \textbf{level of $a$}  & \textbf{$c(a)$} & \textbf{preference list of $a$ }&  & \textbf{$\qquad \ \ c(b)$}  \\
      \hline
      \hline 
      $0, \ldots, \T  - 1$ &  & \preflqa & $q^-(b)$ & \\ [2pt]
      \cline{1-1} \cline{3-5} 
      $\T$, $\T + 1$ & $q^+(a)$ & \prefa & $q^-(b)$ & if $\exists\ a_i\in M(b)$ such that $a_i$ is at level $< \T$ \\ [2pt]
      \cline{1-2}
      $\T+2, \ldots, \S+\T+1$ & $q^-(a)$ &  & $q^+(b)$ & otherwise \\[2pt]
\hline     
\hline
\end{tabular}
\vspace{0.1in}
\caption{Capacities and preference lists used when $a\in\AA$ proposes to $b\in\BB$ in the algorithm of Nasre \etal~\cite{NNRS2024popular}. Here, $c(v)$ denotes the capacity used for vertex $v$, and $\preflqa$ is $\prefa$ restricted to \emph{lq}-vertices.}
\label{tab:quota}
\end{table}

\section{The stable critical matching problem is NP-complete}
\label{subsec:NPhard}

Before considering critical relaxed-stable matchings, we examine the computational difficulty of requiring stability together with criticality. Given an instance $G=(\AA\cup\BB,E)$ of the many-to-many setting with ties and lower quotas on both sides of the bipartition, the \emph{stable critical matching problem} asks whether $G$ admits a matching that is both stable and critical. We show that this problem is NP-complete even in the one-to-one setting with complete preference lists and lower quotas only on the hospital side.

As mentioned earlier, Goko \etal~\cite{GokoMMY22} study the related \HRTMSLQ\ problem, which seeks a stable matching maximizing the total satisfaction ratio of hospital lower quotas. Such a matching need not be critical: criticality requires minimum deficiency among all matchings, without imposing stability. Theorem~13 of~\cite{GokoMMY22} shows that the \HRTMSLQ\ problem is NP-hard even in the one-to-one setting. Our proof adapts the auxiliary-hospital construction used in the proof of that theorem. 

\begin{theorem}\label{cl:nphardCS}
The stable critical matching problem is NP-complete even when all preference lists are complete, every vertex has an upper quota of one, and positive lower quotas occur only on the hospital side. Hardness holds even when ties occur only on one side, with at most one tie of length two per list.
\end{theorem}

\begin{proof}
The problem clearly belongs to NP because stability can be verified in polynomial time, and criticality can be verified by comparing the matching's deficiency with the minimum possible deficiency, which is polynomial-time computable~\cite{Kavitha2021,NNRS2024popular}.

For NP-hardness, we reduce from \textsc{Perfect-SMTI}, which asks whether a stable marriage instance with ties and incomplete lists admits a perfect stable matching. The construction in the proof of Theorem~2 of~\cite{manlove2002hard} establishes hardness when ties occur only on one side, every non-strict list consists entirely of a single tie of length two, and the target cardinality equals the number of vertices on either side. Thus, \textsc{Perfect-SMTI} is NP-complete under these restrictions.

Let $I$ be such an instance on $G=(\RR\cup\HH,E)$, where $|\RR| = |\HH| = n$. We construct $I'$ by adding an auxiliary hospital $h_0$ and making the graph between $\RR$ and $\HH'=\HH\cup\{h_0\}$ complete. Every vertex has upper quota one, and the lower quotas are as follows: 
$q^-(r)=0$ for each $r \in \RR$, $q^-(h) = 1$ for $h \in \HH$, and $q^-(h_0) = 0$. Each resident's list in $I'$ consists of its original list, followed by $h_0$, followed by all originally unacceptable hospitals in an arbitrary fixed strict order. Each original hospital retains its list and appends all originally unacceptable residents in an arbitrary fixed strict order. Finally, $h_0$ ranks all residents in an arbitrary strict order.

This construction preserves all original comparisons and introduces no new ties. Since every original non-strict list consists entirely of a single tie of length two, its completion has exactly one tie, occupying the first two positions. All lists on the other side remain strict. The construction takes polynomial time. The complete bipartite graph of $I'$ admits a matching assigning the $n$ residents to the hospitals in $\HH$, so its minimum deficiency is zero. Since only these hospitals have positive lower quotas, a matching in $I'$ is critical if and only if it matches every hospital in $\HH$. Every such matching also matches all residents and leaves $h_0$ unmatched. We show that $I$ admits a perfect stable matching if and only if $I'$ admits a stable critical matching. 

Suppose first that $M$ is a perfect stable matching in $I$. Clearly, the matching $M$ in $I'$ leaves $h_0$ unmatched. It is critical because it matches every hospital in $\HH$. No pair in $E \setminus M$ blocks $M$ in $I'$, since the original comparisons are preserved. Every resident is assigned to an originally acceptable hospital and strictly prefers it to $h_0$ and to each of the originally unacceptable hospitals. Hence, no newly added pair blocks $M$, so $M$ is stable in $I'$.

Conversely, let $M$ be a stable critical matching in $I'$. It matches all residents and all hospitals in $\HH$, while $h_0$ is unmatched. If $M$ contains an edge $(r,h) \notin E$, then $r$ strictly prefers $h_0$ to $h$. Since $h_0$ is unmatched and finds $r$ acceptable, $(r,h_0)$ blocks $M$, a contradiction. Thus, $M \subseteq E$, and $M$ is a perfect matching in $I$. Any pair blocking $M$ in $I$ would also block it in $I'$, because the original comparisons are preserved. Therefore, $M$ is stable in $I$.

This proves NP-hardness under the stated restrictions and, together with membership in NP, completes the proof.
\end{proof} 
\section{\CRSM\  in the many-to-many setting}\label{sec:Algorithm}

We now turn to relaxed stability and extend this notion to the many-to-many setting with ties and lower quotas on both sides of the bipartition by generalizing Definition~\ref{def:rsmHRLQ}.

\begin{definition}[Relaxed Stability in the Many-to-Many Setting] \label{def:rsmMM}
A matching $M$ is \RSM\ if, for every blocking pair $(a,b)$ with respect to $M$, at least one of the following conditions holds:
\begin{enumerate}[label=C\arabic*., ref=C\arabic*]
    \item \label{itm:1rsm} $a$ is fully subscribed, and $|M(b')| \le q^-(b')$ for every $b'\in M(a)$ such that $b \succ_a b'$.

    \item \label{itm:2rsm} $b$ is fully subscribed, and $|M(a')| \le q^-(a')$ for every $a' \in M(b)$ such that $a\succ_b a'$.
\end{enumerate}
\end{definition}

A blocking pair $(a,b)$ is \emph{justified from the $a$-side} (respectively, \emph{$b$-side}) if condition~\ref{itm:1rsm} (respectively, condition~\ref{itm:2rsm}) holds. The following claim shows that Definition~\ref{def:rsmMM} coincides with the definition of Krishnaa \etal~\cite{krishnaa2023envy} when restricted to the \HRLQ\ setting.

\begin{cl}\label{cl:equvDef}
For an \HRLQ\ instance, a matching $M$ is \RSM\ according to Definition~\ref{def:rsmMM} if and only if it is \RSM\ according to Definition~\ref{def:rsmHRLQ}.
\end{cl}

\begin{proof}
Suppose $M$ satisfies Definition~\ref{def:rsmMM}, and let $(a,b)$ be any blocking pair. If it were justified from the $b$-side, then $b$ would be fully subscribed and, because the pair blocks $M$, would have a partner $a'$ with $a \succ_b a'$. But $|M(a')| = 1 > q^-(a') = 0$, contradicting condition~\ref{itm:2rsm}. Thus, the pair must be justified from the $a$-side. Condition~\ref{itm:1rsm} therefore requires $a$ to be fully subscribed. Since $q^+(a) = 1$, it is matched to a unique hospital $b'$. Blocking implies $b \succ_a b'$, so the same condition gives $|M(b')|\le q^-(b')$. Hence, no unmatched resident blocks $M$, and every blocking resident's hospital is non-surplus, as required by Definition~\ref{def:rsmHRLQ}.

For the converse, suppose $M$ satisfies Definition~\ref{def:rsmHRLQ}. For every blocking pair $(a,b)$, resident $a$ is matched to a hospital $b'$ with $|M(b')| \le q^-(b')$. Since $q^+(a)=1$, $a$ is fully subscribed, and blocking implies $b \succ_a b'$. Thus, condition~\ref{itm:1rsm} holds for every blocking pair, so $M$ satisfies Definition~\ref{def:rsmMM}.
\end{proof}

\subsection{Algorithm}\label{subsec:AlgoMMRSM}

As mentioned earlier, our algorithm combines Kir{\'a}ly's algorithm, adapted to the many-to-many setting, with the popular critical matching algorithm discussed in the previous section.

\subsubsection{Algorithm overview}
Recall that $\S$ and $\T$ denote the sums of lower quotas of vertices in $\AA$ and $\BB$, respectively, \textit{i.e.}, $\S=\sum_{a\in\AA}q^-(a)$ and $\T = \sum_{b\in\BB}q^-(b)$. At each level, vertices in $\AA$ propose, and vertices in $\BB$ accept or reject. At a very high level, the algorithm has three parts, corresponding to different ranges of levels. At levels $0, \ldots, \T-1$, proposals are restricted to \emph{lq}-vertices in $\BB$ and prioritize their lower quotas. This ensures the criticality on the $\BB$ side. At level $\T$ and sub-level $\T^*$, the algorithm uses Kir{\'a}ly's step to handle ties. This removes unjustified blocking pairs and provides the required size guarantee. At levels $\T+1, \ldots, \S+\T$, only deficient vertices in $\AA$ make proposals. This ensures criticality on the $\AA$ side. These parts are interleaved: different vertices may be processed at different levels. If $\T=0$, the first part is empty; if $\S=0$, the third part is empty.

The neighbors to which a vertex $a \in \AA$ proposes, and the order in which it considers them, depend on its level. Let $\prefsa$ denote the strict preference list obtained from $\prefa$ by ordering vertices within each tie by increasing index. Let $\prefslqa$ be the list obtained from $\prefsa$ by deleting all non-\emph{lq} vertices.
For example, suppose $\prefa = (b_2, b_1), b_5, (b_3, b_4)$ where $b_4$ and $b_5$ are \emph{lq} vertices. Here, $a$ ranks
$b_1$ and $b_2$ as rank-1, $b_5$ as rank-2 and $b_3$ and $b_4$ as rank-3. We have $\prefsa = b_1, b_2, b_5, b_3, b_4$ and $\prefslqa = b_5, b_4$ where comma separated vertices denote a strict ordering.

Receivers use their original preference lists, possibly containing ties, together with the proposer's levels. A higher level proposer is preferred over a lower level proposer, irrespective of the original preference ranks. For these comparisons, $\T^*$ has level $\T$. Among partners at level $\T$, original preference ranks are compared first, and starred status breaks ties in favor of the starred partner. The term \emph{least-preferred partner} refers to this ordering. The uncertain-proposal rejection rule, described below, provides an additional rule for selecting a partner to reject.

The pseudocode appears in Algorithm~\ref{algo:maxCRSMM}. Initially, all vertices in $\AA$ enter the queue at level $0$. We write $a^\ell$ when $a$ is processed at level $\ell$. Proposal histories are maintained separately at each level. An edge $(a^\ell,b)\in M$ records the level attached to that particular pair; different edges incident to $a$ may carry different levels. Throughout the algorithm, $|M(a)|$ counts all partners of $a$, irrespective of these labels.

\begin{algorithm}[t]
    \SetKwFunction{decision}{DecideAccRej}
    \SetKwFunction{deficient}{CriticalPropose}
    \SetKwFunction{tiesproposeA}{TiesProposeNormal}
    \SetKwFunction{tiesproposeB}{TiesProposeStar}
    \caption{Critical relaxed-stable matching in $G = (\AA \cup \BB, E)$ }\label{algo:maxCRSMM}
    \DontPrintSemicolon
    \SetAlgoLined
    \begin{spacing}{1.2}
    let $\S=\sum_{a\in\AA}q^-(a)$, $\T = \sum_{b\in\BB}q^-(b) $ and set $M=\emptyset$\;
    \end{spacing}
    initialize a queue $Q=\{a^0\ :\ a\in \AA\}$\;
    \textbf{for} each $a\in \AA$ and $k\in\{\T, \T^*\}$ \textbf{do} $a^k$  unmarks all $b\in \prefa$\;
    \While{$Q$ is not empty}{
        let $a^\ell=dequeue(Q)$\;
        \If {$\ell<\T$ \label{alg:all2lqs} 
        } { 
            \If{ $a^\ell$ has not exhausted $\prefslqa$\label{alg:onlyLQ1} }{let $b=$  most preferred unproposed vertex by $a^\ell$ in $\prefslqa$ \label{alg:onlyLQ2}\;
            \deficient{$a^\ell,q^+(a),b,q^-(b)$} \label{alg:CriticalPropose1}\;
            }
            \Else{add $a^{\ell+1}$ to $Q$ \label{alg:all2lqe}\;}
        }
        \uElseIf{$\ell==\T$ \label{alg:kiralyS1}}{
            \tiesproposeA{$a^\T$}\label{alg:kiralyE}\;
        }
        \uElseIf{$\ell==\T^*$ \label{alg:kiralyS12}}{
            \tiesproposeB{$a^{\T^*}$}\label{alg:kiralyE2}\;
        }
        \Else{ \label{alg:LQAs} 
             \If{$a^\ell$ has not exhausted $\prefsa$}{
                let $b=$ most preferred unproposed vertex by $a^\ell$ in $\prefsa$\;
                   \textbf{if} there exists $a_j^y\in M(b)$ such that $y<\T$ \textbf{then} $c(b)=q^-(b)$ \textbf{else} $c(b)=q^+(b)$\label{alg:capB} \;
                   \deficient{$a^\ell,q^-(a),b,c(b)$}\label{alg:criticalP2} \; }
                \Else{ 
                \If{$\ell<\S+\T$ and $|M(a)|<q^-(a)$}{add $a^{\ell+1}$ to $Q$  \label{alg:LQAe}\;}
                 }
        }
    }
    \Return the underlying matching $M$, ignoring level and starred labels
\end{algorithm}

\subsubsection{Proposals to \textit{lq}-vertices}
 At a level below $\T$ (see Lines~\ref{alg:all2lqs}--\ref{alg:all2lqe} of Algorithm~\ref{algo:maxCRSMM}), an under-subscribed vertex $a$ proposes to vertices in $\prefslqa$. Its \emph{most-preferred} neighbor in $\prefslqa$ or $\prefsa$ is the best-ranked neighbor to which it has not yet proposed at the current level. If $a$ remains under-subscribed after exhausting $\prefslqa$ at level $\ell<\T$, then $a$ raises its level to $\ell + 1$.  It continues using $\prefslqa$ if $\ell + 1 < \T$, and enters Kir{\'a}ly's step otherwise. 
 
In this part, we invoke $\texttt{CriticalPropose}(a^\ell, q^+(a), b, q^-(b))$. Thus, $a$ has its capacity set to $q^+(a)$, while proposals to $b$ use a capacity $q^-(b)$. If $a$ is already matched to $b$ at a lower level, the procedure upgrades the existing edge. Otherwise, $b$ accepts if it has an unfilled position. When full relative to this capacity, $b$ compares the proposal with a least-preferred partner. A vertex $b\in\BB$ prefers $a_i^\ell$  to $a_j^{\ell'}$ if either: (i)  $\ell > \ell'$ or (ii) $\ell = \ell'$ and $a_i\succ_b a_j$.

\subsubsection{Kir{\'a}ly's step}
A vertex that remains under-subscribed after exhausting $\prefslqa$ at level $\T-1$ enters level $\T$. When $\T=0$, vertices start directly at this level. Here, each vertex uses its original preference list $\prefa$, possibly containing ties. The many-to-many adaptation of Kir{\'a}ly's algorithm is implemented by \texttt{TiesProposeNormal()} and \texttt{TiesProposeStar()} (see Lines~\ref{alg:kiralyS1}--\ref{alg:kiralyE2} of Algorithm~\ref{algo:maxCRSMM}). At level $\T$, an existing edge from a lower level is upgraded when its proposer proposes the corresponding receiver. For other proposals, \texttt{TiesProposeNormal()} uses the temporary capacity
\[
c(b)= \begin{cases}
q^-(b), & \text{if $b$ has a partner below level $\T$},\\
q^+(b), & \text{otherwise}.
\end{cases}
\]
If $b$ has an unfilled position relative to $c(b)$, it accepts the proposal. Otherwise, it first rejects a least-preferred partner below level $\T$, if one exists. If no such partner exists, it rejects a partner whose level-$\T$ proposal is \emph{uncertain}, if one exists. In the remaining case, it accepts the proposal only if its least-preferred partner has level $\T$ or $\T^*$ and is strictly worse than the proposer in $\prefb$. In particular, a level-$\T$ proposal cannot displace a partner at a level above $\T$.

We use level-specific versions of the favorite-neighbor and uncertain-proposal notions from Section~\ref{sec:GenKiraly}. In these notions, under-subscription is always measured relative to the original upper quota, rather than the temporary capacity.

A first proposal from $a^\T$ to a rank-$k$ neighbor $b$, if accepted, is labeled uncertain if another rank-$k$ neighbor $b''\ne b$ is unproposed by $a^\T$ and satisfies $|M(b'')|<q^+(b'')$. Here, ``first'' refers to the first proposal at level $\T$, irrespective of proposals at earlier levels. The test is applied after every accepted first proposal at level $\T$, including an upgrade of an existing edge or acceptance by rejecting a lower-level partner (see Lines~\ref{tiesP:uncertain1}--\ref{tiesP:uncertain2} of \texttt{TiesProposeNormal()}). When $b$ rejects $a^{\T}$ because of the uncertain-proposal, the vertex $a^\T$ marks $b$ for a later proposal (see Lines~\ref{tiesP:uncertain3}--\ref{tiesP:uncertain4}).
 
\begin{definition}
[Favorite Neighbor of $a^\ell$ for $\ell\in\{\T,\T^*\}$] \label{def:favNbrLvl}
Let $k$ be the best rank in $\prefa$ containing a neighbor that is unproposed or marked by $a^\ell$. The favorite neighbor $b$ of $a^\ell$ is selected as follows:
\begin{enumerate}[label=(\roman*)]
    \item \label{itm:FavNbrT1} If there is an under-subscribed neighbor at rank $k$ that is unproposed by $a^\ell$, choose the one with the smallest index.

    \item \label{itm:FavNbrT2} Otherwise, if there is a fully subscribed neighbor at rank $k$ that is unproposed by $a^\ell$, choose the one with the smallest index.

    \item \label{itm:FavNbrT3} Otherwise, choose the smallest-index neighbor marked by $a^\ell$ at rank $k$.
\end{enumerate}
\end{definition}

If $a$ remains under-subscribed after processing all unproposed and marked neighbors at level $\T$, it attains starred status and enters sub-level $\T^*$ (see Line~\ref{tiesP:Tstar} of \texttt{TiesProposeNormal()}). Before this transition, all its remaining uncertain proposals are labeled `not uncertain' (see Line~\ref{TiesNormal:notUncertain}). No new uncertainty labels are assigned at sub-level $\T^*$. The interpretation of starred status is the same as in Section~\ref{sec:GenKiraly}. At sub-level $\T^*$, the vertex reproposes its neighbors using \texttt{TiesProposeStar()}. If it is already matched to the selected neighbor $b$ at level at most $\T$, the existing edge is upgraded to level $\T^*$. Otherwise, let $a_w^\ell$ be a least-preferred partner of $b$ under the ordering described above. The vertex $b$ rejects $a_w^\ell$ precisely when $a$ and $a_w$ are tied in $\prefb$ and $\ell$ is a level at most $\T$. In all other cases, $b$ rejects the proposal made by $a$ (see Lines~\ref{TiesPropose1:atSProp1}--
\ref{TiesPropose1:atSProp3} of \texttt{TiesProposeStar()}).

\begin{proced}
\caption{\tt CriticalPropose($a^\ell,q_a,b,q_b$)}
    \DontPrintSemicolon
    \SetAlgoLined
    \If{$a^x\in M(b)$ for $x<\ell$ \label{dec:art1}}{$M=(M\setminus \{(a^x,b)\}) \cup \{(a^\ell,b)\}$\label{dec:art2}}
    \uElseIf{$|M(b)|<q_b$\label{dec:st1}}{$M=M\cup \{(a^\ell,b)\}$\;}
    \uElseIf{$|M(b)|==q_b$}{
        let $a_j^y\in M(b)$ be the least preferred vertex by $b$ in $M(b)$\;
        \If{($\ell>y$) or ($\ell==y$ and $a\succ_b a_j$ )}{
        $M=M\setminus \{(a_j^y,b)\}\cup \{(a^\ell,b)\}$ and add $a_j^y$ to $Q$ if $\forall x \ a_j^x\notin Q$\;}\label{dec:st2}}
        \If{ $|M(a)|<q_a $ and  $\forall x\  a^x\notin Q$\label{dec:que1}}  {add $a^\ell$ to $Q$ \label{dec:que2}
        }  
\end{proced}

\subsubsection{Proposals by \textit{lq}-vertices}
If $a$ remains deficient\footnote{In the algorithm considered by Nasre \etal~\cite{NNRS2024popular}, a vertex advances to level $\T+1$ if it remains under-subscribed. In our algorithm, it advances to level $\T+1$ only if it remains deficient. Our levels $\T+1,\ldots,\S+\T$ correspond to levels $\T+2,\ldots,\S+\T+1$ in their algorithm.} after exhausting $\prefa$ at sub-level $\T^*$, it advances to level $\T+1$ (see Line~\ref{tiesP:deficient} of \texttt{TiesProposeStar()}). It then proposes using $\prefsa$, which contains all its neighbors, including non-\emph{lq} vertices (see Lines~\ref{alg:LQAs}--\ref{alg:LQAe} of Algorithm~\ref{algo:maxCRSMM}).

At these levels, the algorithm invokes $\texttt{CriticalPropose}(a^\ell,q^-(a),b,c(b))$, where $c(b)$ is defined as above. A deficient vertex that exhausts its list at a level below $\S+\T$ advances to the next level. It stops proposing when it becomes non-deficient or exhausts its list at level $\S+\T$. A vertex that becomes non-deficient resumes proposing if a subsequent rejection makes it deficient again. Similarly, a vertex that is non-deficient after exhausting sub-level $\T^*$ makes no further proposals unless it subsequently becomes deficient.

\subsubsection{Termination and proposal bound}
The algorithm terminates when the queue $Q$ is empty. At termination, every deficient vertex in $\AA$ has exhausted $\prefsa$ at level $\S+\T$, and every non-deficient under-subscribed vertex in $\AA$ has exhausted $\prefa$ at sub-level $\T^*$. We note that $\S+\T = O(m)$, where $m = |E|$ (it is natural to assume that the upper quota $q^+(u)$ for the vertex $u$ is at most the total number of neighbors of $u$ in $G$). Each edge is processed at most once at every level other than $\T$. At level $\T$, it is processed at most twice: once as an unproposed edge and at most once more after being marked. At sub-level $\T^*$, each edge is processed at most once. Thus, each edge is processed at most $\S+\T+3$ times and hence, the total number of proposals is $O(m^2)$.

\begin{proced}[t]
    \caption{\tt TiesProposeNormal($a^\T$)}
    \label{proce:TiesProp}
    \DontPrintSemicolon
    \SetAlgoLined
    \If{there exists a vertex in $\prefa$ which is marked or unproposed by $a^\T$}{
        let $b$ be the favorite neighbor of $a^\T$ in $\prefa$ at some rank, say $k$ \label{tiesP:favnbr1}\;
        label uncertain proposals from $a^\T$, if any, at rank $k-1$ as `not uncertain'\;
        let $\mathit{first}$ be true if $b$ is unproposed by $a^\T$, and false otherwise\;
        \textbf{if} $b$ was marked by $a^\T$ \textbf{then} $a^\T$ unmarks $b$\;
        $a^\T$ proposes to $b$\;

        \eIf{$a^x\in M(b)$ for some $x<\T$}{
            $M=(M\setminus\{(a^x,b)\})\cup\{(a^\T,b)\}$\;
        }{
            \textbf{if} $\exists\,a_j^y\in M(b)$ such that $y<\T$
            \textbf{then} $c(b)=q^-(b)$
            \textbf{else} $c(b)=q^+(b)$ \label{tiesP:CapB}\;

            \eIf{$|M(b)|<c(b)$}{
                $M=M\cup\{(a^\T,b)\}$\;
            }{
                \uIf{$\exists\,a_i^x\in M(b)$ such that $x<\T$}{
                    let $a_j^y\in M(b)$ be a least-preferred partner in $M(b)$\;
                    $M=(M\setminus\{(a_j^y,b)\}) \cup\{(a^\T,b)\}$\;
                    add $a_j^y$ to $Q$ if $\forall z\;a_j^z\notin Q$\;
                }
                \uElseIf{$\exists\,a_w^\T\in M(b)$ such that the proposal $(a_w^\T,b)$ is uncertain \label{tiesP:uncertain3}}{
                    $M=(M\setminus\{(a_w^\T,b)\}) \cup\{(a^\T,b)\}$\;
                    $a_w^\T$ marks $b$\;
                    add $a_w^\T$ to $Q$ if $\forall x\;a_w^x\notin Q$ \label{tiesP:uncertain4}\;
                }
                \Else{
                    let $a_w^\ell$ be a least-preferred partner in $M(b)$\;
                    \If{$\ell\in\{\T,\T^*\}$ and $a\succ_b a_w$}{
                        $M=(M\setminus\{(a_w^\ell,b)\}) \cup\{(a^\T,b)\}$\;
                        add $a_w^\ell$ to $Q$ if $\forall x\;a_w^x\notin Q$\;
                    }
                }
            }
        }

        \If{$\mathit{first}$ and $(a^\T,b)\in M$}{
            \If{$\exists\,b''\neq b$ at rank $k$ in $\prefa$ with $|M(b'')|<q^+(b'')$ and $b''$ is unproposed by $a^\T$ \label{tiesP:uncertain1}}{
                label $(a^\T,b)$ as an uncertain proposal \label{tiesP:uncertain2}\;
            }
        }

        \textbf{if} $|M(a)|<q^+(a)$ \textbf{then} add $a^\T$ to $Q$ if $\forall z\;a^z\notin Q$ \label{alg:underSubA}\;
    }
    \Else{
        add $a^{\T^*}$ to $Q$ if $\forall x\ a^x\notin Q$ \label{tiesP:Tstar}\;
        label uncertain proposals from $a^\T$, if any, as `not uncertain' \label{TiesNormal:notUncertain}\;
    }
\end{proced}

\begin{proced}[t]
    \caption{\tt TiesProposeStar($a^{\T^*}$)}\label{proce:TiesPropT1}
    \DontPrintSemicolon
    \SetAlgoLined
  \If{$a^{\T^*}$ has not proposed to all vertices in $\prefa$}
        {   let $b$ be the favorite neighbor of $a^{\T^*}$\label{tiesP:favnbr2}\;
         $a^{\T^*}$ proposes to $b$\label{TiesPropose1:atSProp}\;
            \If{$(a^x\in M(b)$ for $x\le\T)$}
            {   $M=(M\setminus \{(a^x,b)\}) \cup \{(a^{\T^*},b)\}$
            }
            \Else
            {  let $a_w^\ell$ be one of the least preferred partners in $M(b)$\label{TiesPropose1:atSProp1}\;
                \If{$b$ ranks both $a$ and $a_w$ at the same rank and $\ell \le \T$ and $\ell \neq \T^*$ \label{TiesPropose1:atSProp2}}
                {   $M=(M\setminus \{(a_w^{\ell},b)\}) \cup \{(a^{\T^*},b)\}$ and add $a_w^{\ell}$ to $Q$ if $\forall z\  a_w^z\notin Q$\label{TiesPropose1:atSProp3}                
                }
            }
            \textbf{if} $|M(a)|< q^+(a)$ \textbf{then} add $a^{\T^*}$ to $Q$ if $\forall z\  a^z\notin Q$\;
        }
        \textbf{else if} $|M(a)|<q^-(a)$ \textbf{then} add $a^{\T+1}$ to $Q$ if $\forall z\  a^z\notin Q$ \label{tiesP:deficient}\;    
\end{proced}

Before proving correctness, we establish some invariants of the algorithm that will be used throughout the analysis. In the following observation, $M$ denotes the current matching during execution.

\begin{observation}\label{obs:algorithm-invariants}
For every $b\in\BB$, the following hold:
\begin{enumerate}[label=(\roman*), ref=\roman*]
    \item \label{obs:degree-monotonicity}
    The number of matched partners of $b$ never decreases.

    \item \label{obs:temporary-capacity}
    If $b$ is matched to a partner below level $\T$, then $|M(b)|\le q^-(b)$.

    \item \label{obs:normal-rejection}
    If some level-$\T$ vertex, say $a^{\T}$, is rejected by $b$, then $b$ is fully subscribed relative to $q^+(b)$ and all vertices in $M(b)$ have levels at least $\T$. These two conditions continue to hold thereafter.
\end{enumerate}
\end{observation}

\begin{proof}
\noindent\textbf{Proof of \ref{obs:degree-monotonicity}:} Every update adds an edge, replaces a partner, or changes an edge's level. So, the number of matched partners never decreases. \hfill $\triangleleft$

\noindent\textbf{Proof of \ref{obs:temporary-capacity}:} 
We argue inductively over the updates to $M$. Initially, the claim holds since $M=\emptyset$.

Suppose $b$ already has a partner below level $\T$. Proposals below level $\T$ use capacity $q^-(b)$ in \texttt{CriticalPropose()} (see Line~\ref{alg:CriticalPropose1} of Algorithm~\ref{algo:maxCRSMM}), while proposals at level $\T$ or above use the same capacity whenever such a partner is present (see Line~\ref{tiesP:CapB} of \texttt{TiesProposeNormal($a^\T$)}, and Line~\ref{alg:capB} of Algorithm~\ref{algo:maxCRSMM}). Hence, a new partner is added only when $|M(b)| < q^-(b)$. All other updates, including those in \texttt{TiesProposeStar()}, preserve $|M(b)|$. Thus, the bound is maintained.

If $b$ had no partner below level $\T$, it can be matched to one only by accepting a proposal below level $\T$. Such a proposal uses \texttt{CriticalPropose()} with $q_b = q^-(b)$. An addition requires $|M(b)| < q^-(b)$, and a replacement requires $|M(b)| = q^-(b)$ (see Lines~\ref{dec:st1}--\ref{dec:st2} of \texttt{CriticalPropose()}). Consequently, the bound holds. \hfill $\triangleleft$

\noindent\textbf{Proof of \ref{obs:normal-rejection}:} A level-$\T$ proposal to an under-subscribed $b$ is always accepted, possibly by upgrading an existing edge or replacing a lower-level partner. Moreover, neither \texttt{CriticalPropose()} nor \texttt{TiesProposeNormal()} can reject a matched partner at level at least $\T$ while $b$ is under-subscribed. This is because a proposal below level $\T$ cannot reject a partner at level at least $\T$; a proposal at level at least $\T$ can reject such a partner only when no lower-level partner is in $M(b)$, and $|M(b)| = q^+(b)$. 

Suppose, in \texttt{TiesProposeStar()}, a proposal from $a^{\T^*}$ causes the first rejection at $b$ of a partner at level at least $\T$ while $b$ is under-subscribed. By part~\ref{obs:degree-monotonicity}, $b$ was also under-subscribed when $a$ previously proposed to it at level $\T$. That proposal was accepted, and the original pair $(a,b)$ could not have been removed before this first such rejection. Hence, \texttt{TiesProposeStar()} would upgrade the existing edge, a contradiction.

Therefore, rejection of $a^\T$ requires $b$ to be fully subscribed relative to $q^+(b)$. No partner below level $\T$ can be in $M(b)$ at the time of such a rejection since lower-level partners are prioritized for rejection. By part \ref{obs:degree-monotonicity} above, $b$ continues to remain fully subscribed with respect to $q^+(b)$. Once $b$ is fully subscribed with all partners at levels at least $\T$, $b$ cannot subsequently accept a proposal below level $\T$. 
\end{proof}

\begin{observation}\label{obs:uncertainty-invariants}
During the execution of the algorithm, the following hold for every $b \in \BB$:
\begin{enumerate}[label=(\roman*)]
    \item \label{obs:no-new-uncertainty} If $b$ is fully subscribed before receiving a proposal, that proposal cannot be labeled uncertain.

    \item \label{obs:advanced-proposals} While $b$ holds an uncertain proposal, every proposal to $b$ at sub-level $\T^*$ or a level above $\T$ comes from a vertex already matched to $b$.
\end{enumerate}
\end{observation}

\begin{proof}
    \noindent\textbf{Proof of \ref{obs:no-new-uncertainty}:} Suppose, for a contradiction, that the incoming proposal is accepted and labeled uncertain. Then the proposer has another unproposed, under-subscribed neighbor tied with $b$. Since $b$ was fully subscribed before receiving the proposal, the favorite-neighbor rule would have selected such a neighbor instead, a contradiction. Since the number of partners of $b$ never decreases, once $b$ is fully subscribed, subsequent proposals to $b$ cannot be labeled uncertain. \hfill $\triangleleft$

    \noindent\textbf{Proof of \ref{obs:advanced-proposals}:} Suppose, for a contradiction, that $b$ receives a proposal at sub-level $\T^*$ or a level above $\T$ from a vertex not matched to $b$, while $b$ holds an uncertain proposal. Consider the first such proposal, and let $a'$ be its proposer. Since $a'$ has exhausted level $\T$, its last level-$\T$ proposal to $b$ was rejected immediately or subsequently. This implies that $b$ was fully subscribed at the time of rejection. Hence, by part~\ref{obs:no-new-uncertainty}, the uncertain proposal currently held by $b$ was already present.
    
    An immediate rejection of the last level-$\T$ proposal from $a'$ to $b$ is therefore impossible, since $b$ already held an uncertain proposal and would accept the incoming proposal, rejecting a lower-level partner or an uncertain partner. Since $a'$ clears its uncertain labels on leaving level $\T$, an uncertainty-based rejection would occur before this transition and force a marked reproposal, contradicting the choice of its last level-$\T$ proposal. A proposal below level $\T$ cannot cause the rejection of $a'$, and a level-$\T$ proposal cannot cause its rejection by any other rule while $b$ holds an uncertain proposal. Thus, the rejection must have resulted from an earlier starred or higher-level proposal from a non-partner while $b$ held an uncertain proposal, contradicting our choice.
\end{proof}

\subsection{Correctness}\label{subsec:correctnessM1}

We prove the following properties of the matching $M$ computed by our algorithm:
\begin{enumerate}[label=\Roman*.]
    \item $M$ is critical and relaxed-stable (\RSM).
    \item $M$ has size at least $\frac{2}{3}$ of that of a maximum-size relaxed-stable matching in $G$, whether or not it is critical.
\end{enumerate}

To establish criticality and relaxed stability, we construct a \emph{cloned} graph $G_M$. The construction is similar to those in~\cite{brandl2019two,NNRS2024popular}. For each vertex $u$, we number its $q^+(u)$ positions from $1$ to $q^+(u)$ and designate the first $q^-(u)$ positions as critical. Throughout the paper, we use the convention that if $u$ is matched to $k$ neighbors, its first $k$ positions are filled and its remaining $q^+(u)-k$ positions are unfilled.

\subsubsection{Construction of the cloned graph $G_M$}\label{sec:cloneG}
We construct $G_M$ from the matching $M$ computed by our algorithm. This construction allows us to work with a one-to-one matching $M^*$ corresponding to $M$.

\noindent\textbf{Vertex set of $G_M$:}
For every vertex $v \in \AA \cup \BB$, we create $q^+(v)$ \emph{clones}, denoted by  $v_1,\ldots,v_{q^+(v)}$. The vertex set of $G_M$ is $\AA'\cup\BB'$, where 
$$
\AA' = \{ \ a_i :  a \in \AA \ \mbox{and } 1 \le i \le q^+(a)\} \hspace{0.25in} \text{ and } \hspace{0.25in}
\BB' = \{ \ b_j :  b \in \BB \  \mbox{and } 1 \le j \le q^+(b)\}\\
$$
The clone $v_i$ represents the $i$th position of $v$. Accordingly, clones $v_i$ with $i \le q^-(v)$ are called \emph{lower-quota clones} or \emph{critical clones}, and the remaining clones are called \emph{non-critical clones}.

\noindent\textbf{Edge set of $G_M$:}
The edge set is $E'= M^* \cup E'_U$, where $M^*$ is the one-to-one matching representing $M$, and $E'_U$ consists of the edges outside $M^*$. We construct these sets as follows:
\begin{enumerate}
    \item \textbf{The matching $M^*$:} Initially, set $M^* = \emptyset$. Process the edges of $M$ in an arbitrary fixed order. For each $(a,b)\in M$, let $a_i$ and $b_j$ be the lowest-index unmatched clones of $a$ and $b$, respectively, and add $(a_i, b_j)$ to $M^*$. Such clones exist because $M$ respects the upper quotas. Only this representative edge is included for the pair $(a,b)\in M$.

    \item \textbf{The unmatched edges $E'_U$:} \label{itm:unmatched}
    For every $(a,b)\in E\setminus M$, include in $E'_U$ every edge $(a_i, b_j)$ with $1 \le i \le q^+(a)$ and $1 \le j \le q^+(b)$. Thus, these edges form a complete bipartite graph between the clones of $a$ and the clones of $b$.
\end{enumerate}

By construction, $|M^*| = |M|$, and only the first $|M(v)|$ clones of each vertex $v$ are matched in $M^*$.

We next partition $\AA' \cup \BB'$ and establish properties of these partitions that will be used to prove criticality and relaxed stability of $M$.

\noindent \textbf{Partition of vertices:} 
We partition the clone sets as $\AA' = \bigcup_{x = 0}^{\S + \T} \AA_x$ and $\BB' = \bigcup_{x = 0}^{\S + \T} \BB_x$. The clones are assigned to these partitions as follows:

\begin{itemize}
    \item \label{itm:part111} \textbf{Matched clones in $\AA' \cup \BB'$:} Let $(a_i,b_j)\in M^*$ represent an edge $(a^\ell, b)\in M$, where $\ell \in \{0, \ldots, \S + \T\} \cup \{\T^*\}$. Set $x = \T$ if $\ell = \T^*$, and $x = \ell$ otherwise. Assign $a_i$ to $\AA_x$ and $b_j$ to $\BB_x$.

    \item \label{itm:part411} \textbf{Unmatched clones in $\AA'$:} Let $a_i$ be a clone of $a\in\AA$ that is unmatched in $M^*$. If $i\le q^-(a)$, assign $a_i$ to $\AA_{\S+\T}$; otherwise, assign it to $\AA_{\T}$. Thus, unmatched critical clones are placed in $\AA_{\S + \T}$, and unmatched non-critical clones are placed in $\AA_{\T}$.
    
    \item \label{itm:part5} \textbf{Unmatched clones in $\BB'$:} Let $b_j$ be a clone of $b\in\BB$ that is unmatched in $M^*$. If $j \le q^-(b)$, assign $b_j$ to $\BB_0$; otherwise, assign it to $\BB_{\T}$. Thus, unmatched critical clones are placed in $\BB_0$, and unmatched non-critical clones are placed in $\BB_{\T}$.
\end{itemize}

We draw the partitions from top to bottom in decreasing order of their indices, with $\AA_x$ and $\BB_x$ at the same height; see Figure~\ref{fig:maxlevelgraphMM}. To describe edge directions, we regard every edge as oriented from $\AA'$ to $\BB'$. An edge from $\AA_x$ to $\BB_y$ is horizontal if $x = y$, upward if $x<y$, and downward if $x>y$. By construction, every edge of $M^*$ (shown in blue) is horizontal, whereas edges of $E'_U$ (shown in solid black) may be horizontal, upward, or downward. Property~\ref{property:GM} states the properties of the vertices and edges of $G_M$ under this partition.

\begin{property}\label{property:GM}
Let $a\in\AA$ and $b\in\BB$. Then the following hold in graph $G_M$.
\begin{enumerate}
    \item \label{obs:partitionPorp1} The set $\bigcup_{x=\T+1}^{\T+\S}\AA_x$ contains at most $q^-(a)$  clones of $a$. Thus, $|\bigcup_{x=\T+1}^{\T+\S}\AA_x|\le \S$.
    \item \label{obs:partitionPorp2} The set $\bigcup_{x=0}^{\T-1}\BB_x$ contains at most $q^-(b)$  clones of $b$. Thus, $|\bigcup_{x=0}^{\T-1}\BB_x|\le \T$.
    \item \label{obs:partitionPorp3} If $a$ is deficient in $M$, then exactly $q^-(a)$ clones of $a$ belong to $\AA_{\S+\T}$, and its remaining clones belong to $\AA_{\T}$. Moreover, every neighbor in $G_M$ of any clone of $a$ is matched in $M^*$ and belongs to $\BB_{\S+\T}$.
    \item \label{obs:partitionPorp4} If $a$ is under-subscribed in $M$, then all its clones belong to $\bigcup_{x=\T}^{\S+\T}\AA_x$. Moreover, every neighbor in $G_M$ of any clone of $a$ is matched in $M^*$ and belongs to $\bigcup_{x=\T}^{\S+\T}\BB_x$.  
    \item \label{obs:partitionPorp5} If $b$ is deficient in $M$, then all clones of every neighbor of $b$ in $G$ that is not matched to $b$ belong to $\AA_0$.
    \item \label{obs:partitionPorp6} If $b$ is under-subscribed in $M$, then all clones of every neighbor of $b$ in $G$ that is not matched to $b$ belong to $\bigcup_{x=0}^{\T}\AA_x$.
\end{enumerate}
\end{property}

\begin{figure}[t]
\begin{center}
    \scalebox{0.85}{	\begin{tikzpicture}[scale=0.6, thick,fsnode/.style={draw,circle,fill=black,scale=0.4}, lnode/.style={draw=red,circle,fill=red,scale=0.4}, snode/.style={draw=blue,circle,fill=blue,scale=0.4},tnode/.style={draw=black,circle,scale=0.4}, ltnode/.style={draw=red,circle,scale=0.4}]
	\draw[very thick] (1,1.5) ellipse (1cm and 5cm);
	\node at (1,-4) {$\AA'$};
	\draw[very thick] (0.25,5) -- (1.75,5);
	\node at (1,4) {\vdots};
	\draw[very thick] (0.07,3.4) -- (1.93,3.4);
	\draw[very thick] (0,2) -- (2,2);
	\draw[very thick] (0,1) -- (2,1);
	\node at (1,-1) {\vdots};
	\draw[very thick] (0.05,-0.5) -- (1.95,-0.5);
	\draw[very thick] (0.3,-2) -- (1.7,-2);
	\node at (-1.5,5.5) {$\AA_{\S+\T}$};
	\node at (-1.5,2.75) {$\AA_{\T+1}$};
	\node at (-1.5,1.5) {$\AA_{\T}$};
	\node at (-1.5,0.25) {$\AA_{\T-1}$};
	\node at (-1.5,-1) {\vdots};
	\node at (-1.5,-2.5) {$\AA_{0}$};
	\draw[very thick] (8,1.5) ellipse (1cm and 5cm);
	\node at (8,-4) {$\BB'$};
	\draw[very thick] (7.25,5) -- (8.75,5);
	\node at (8,4) {\vdots};
	\draw[very thick] (7.07,3.4) -- (8.93,3.4);
	\draw[very thick] (7,2) -- (9,2);
	\draw[very thick] (7,1) -- (9,1);
	\node at (8,-1) {\vdots};
	\draw[very thick] (7.05,-0.5) -- (8.95,-0.5);
	\draw[very thick] (7.3,-2) -- (8.7,-2);
	\node at (10.5,5.5) {$\BB_{\S+\T}$};
	\node at (10.5,2.75) {$\BB_{\T+1}$};
	\node at (10.5,1.5) {$\BB_{\T}$};
	\node at (10.5,0.25) {$\BB_{\T-1}$};
	\node at (10.5,-1) {\vdots};
	\node at (10.5,-2.5) {$\BB_0$};
	\node[fsnode] (b3) at (8,5.5) {};
	\node[lnode] (a4) at (1,5.5) {};
	\draw[very thick,blue] (a4) -- (b3);
	\node[lnode] (a5) at (1,4.6) {};
	\node[fsnode] (b4) at (8,4.6) {};
    \draw[very thick,blue] (a5) -- (b4);
    \node[fsnode] (a2) at (1,0.6) {};
	\node[lnode] (b2) at (8,0.6) {};
	\draw[very thick,blue] (a2) -- (b2);
    \node[lnode] (a10) at (1,0) {};
	\node[lnode] (b10) at (8,0) {};
	\draw[very thick,blue] (a10) -- (b10);
    \node[lnode] (a9) at (1,2.9) {};
	\node[fsnode] (b9) at (8,2.9) {};
    \draw[very thick,blue] (a9) -- (b9);
    \node[lnode] (a11) at (1,2.3) {};
	\node[lnode] (b11) at (8,2.3) {};
    \draw[very thick,blue] (a11) -- (b11);
	\node[fsnode] (a6) at (1,1.3) {};
	\node[fsnode] (b5) at (8,1.3) {};
	\draw[very thick,blue] (a6) -- (b5);
	\node[fsnode] (a8) at (1,-2.5) {};
	\node[lnode] (b8) at (8,-2.5) {};
	\draw[very thick,blue] (a8) -- (b8);
	\node[ltnode] (b6) at (8,-3) {};
	\node[ltnode] (a1) at (1,6) {};
	\node[tnode] (b7) at (8,1.8) {};
	\node[tnode] (a7) at (1,1.8) {};
	\draw[ultra thin] (a4) -- (b4);
	\draw[ultra thin] (a6) -- (b3);
	\draw[ultra thin] (a6) -- (b7);
	\draw[ultra thin] (a7) -- (b5);
	\draw[ultra thin] (a8) -- (b6);
	\draw[ultra thin] (a1) -- (b3);
    \draw[ultra thin] (a8) -- (b7);
    \draw[ultra thin] (a9) -- (b5);
	

    
	\end{tikzpicture}}
\end{center}
\caption{The graph $G_M$ after re-arranging the vertices based on their levels in $M$. Red nodes represent critical clones, and black nodes represent non-critical clones. Matched clones in $M^*$ are represented by solid nodes, and unmatched vertices are represented by hollow nodes. The blue horizontal lines represent matched edges in $M^*$. Solid black lines represent edges which are not matched in $M^*$. Note that no edge in $G_M$ is of the form $\AA_x\times \BB_y$ for $y\le x-2$.
}
\label{fig:maxlevelgraphMM}
\end{figure}

\begin{proof} We justify all of the above properties as follows:

\noindent\textbf{Proof of \ref{obs:partitionPorp1}:} A vertex $a\in\AA$ reaches a level at least $\T+1$ only when deficient, after exhausting $\prefa$ at sub-level $\T^*$ (see Line~\ref{tiesP:deficient} of \texttt{TiesProposeStar()}). Subsequent proposals use \texttt{CriticalPropose()} with capacity $q^-(a)$ (see Line~\ref{alg:criticalP2} of Algorithm~\ref{algo:maxCRSMM}). Hence, $|M(a)|\le q^-(a)$ thereafter. In the final matching, at most $|M(a)|$ matched clones and $q^-(a)-|M(a)|$ unmatched critical clones of $a$ lie at levels at least $\T+1$, giving at most $q^-(a)$ such clones. A vertex that never reaches a level $\T +1$ is non-deficient at termination and has no clones above $\T$. Summing over all $a\in\AA$ gives the bound $\S$. \hfill $\triangleleft$

\noindent\textbf{Proof of \ref{obs:partitionPorp2}:} If $b$ has a matched partner below level $\T$, then by Observation~\ref{obs:algorithm-invariants}(\ref{obs:temporary-capacity}), we have $|M(b)|\le q^-(b)$. Thus, at most $|M(b)|$ matched clones and $q^-(b) - |M(b)|$ unmatched critical clones of $b$ lie below $\T$, giving at most $q^-(b)$ such clones. If no matched clone of $b$ lies below $\T$, only its unmatched critical clones can lie there, so the same bound holds. Summing over all $b\in\BB$ gives the bound $\T$. \hfill $\triangleleft$

\noindent\textbf{Proof of \ref{obs:partitionPorp3}:} If $a$ remains deficient in $M$, it must have exhausted its preference list at level $\S+\T$. Every edge of $M$ incident to $a$ therefore has this label, since \texttt{CriticalPropose()} upgrades existing edges. Thus, its $|M(a)|$ matched clones and its $q^-(a)-|M(a)|$ unmatched critical clones all belong to $\AA_{\S+\T}$, giving exactly $q^-(a)$ such clones. Its remaining clones are unmatched and non-critical, so they belong to $\AA_{\T}$.

For any $(a,b)\in E\setminus M$, the proposal from $a^{\S+\T}$ was rejected either immediately or subsequently, which implies $b$ is fully subscribed relative to $q^+(b)$ with all partners at level $\S+\T$. Hence, every clone of $b$ is matched and belongs to $\BB_{\S+\T}$. For $(a,b)\in M$, only its representative edge appears in $G_M$, and its endpoints belong to $\AA_{\S+\T}$ and $\BB_{\S+\T}$. Therefore, every neighbor in $G_M$ of any clone of $a$ is matched and belongs to $\BB_{\S+\T}$.
\hfill $\triangleleft$

\noindent\textbf{Proof of \ref{obs:partitionPorp4}:}
If $a$ remains under-subscribed in $M$, it must have exhausted $\prefa$ at level $\T$, more specifically sub-level $\T^*$. Every edge incident to $a$ therefore has a label at least $\T$. Its unmatched clones belong to $\AA_{\T}$ or $\AA_{\S+\T}$. Thus, all clones of $a$ lie at or above $\T$.

For any $(a,b)\in E\setminus M$, the level-$\T$ proposal was rejected immediately or subsequently, which by Observation~\ref{obs:algorithm-invariants}
(\ref{obs:normal-rejection}), implies that $b$ is fully subscribed relative to $q^+(b)$ with all partners at levels at least $\T$. Hence, every clone of $b$ is matched and lies at or above $\T$. For $(a,b)\in M$, only its representative matched edge appears in $G_M$, and both endpoints lie at or above $\T$. Hence, every neighbor in $G_M$ of any
clone of $a$ is matched and lies at or above $\T$. \hfill $\triangleleft$

 \noindent\textbf{Proof of \ref{obs:partitionPorp5}:} Since $b$ is deficient in the final matching and its number of partners never decreases (Observation~\ref{obs:algorithm-invariants}(\ref{obs:degree-monotonicity})), it has fewer than $q^-(b)$ partners throughout the execution. Both \texttt{CriticalPropose()} and \texttt{TiesProposeNormal()} use a capacity at least $q^-(b)$, so $b$ accepts new proposals without rejecting existing partners. Any rejection in  \texttt{TiesProposeStar()} is also impossible, except for an upgrade. Thus, every level-$0$ proposal to $b$ is accepted, and its original pair remains in the final matching.

For any $(a,b)\in E\setminus M$, $a$ therefore never proposed to $b$ at level $0$. Since $b$ is an \emph{lq}-vertex, $a$ has not exhausted $\prefslqa$ and has never advanced beyond level $0$. It must be fully subscribed at termination; otherwise, it would continue proposing. Consequently, all its clones are matched through edges labeled $0$ and belong to $\AA_0$.
\hfill $\triangleleft$

 \noindent\textbf{Proof of \ref{obs:partitionPorp6}:} Let $(a,b)\in E\setminus M$, where $b$ is under-subscribed. Clearly, $a$ never proposed to $b$ at level $\T$; otherwise, its immediate or subsequent rejection would force $b$ to be fully subscribed in $M$. Thus, $a$ has not exhausted level $\T$ and has never advanced beyond it; this further implies that $a$ is fully subscribed at termination. Hence, all its clones are matched and belong to $\bigcup_{x=0}^{\T} \AA_x$.
\end{proof}

For an edge $(a_i,b_j) \in E'$ with $a_i \in \AA_x$ and $b_j \in \BB_y$, we say that it is of the form $\AA_x \times \BB_y$. Such an edge is \emph{steep downward} if $x > y+1$. We next show that $G_M$ contains no such edge.

\begin{cl}\label{cl:noedgeMM}
Let $(a,b) \in E \setminus M$. If some clone $a_i$ of $a$ belongs to $\AA_x$, where $x \ge 1$, then every clone of $b$ belongs to $\bigcup_{y=x-1}^{\S+\T}\BB_y$.
\end{cl}

\begin{proof}
Suppose, for a contradiction, that some clone $b_j$ of $b$ belongs to $\BB_y$ with $y < x-1$. Suppose $a_i$ is unmatched. If $a_i$ is critical, then $a$ is deficient and $x = \S + \T$. Property~\ref{property:GM}(\ref{obs:partitionPorp3}) therefore gives $y = \S + \T = x$. Otherwise, $a_i$ is non-critical, so $x = \T$, and $a$ is under-subscribed. Property~\ref{property:GM}(\ref{obs:partitionPorp4}) then gives $y \ge \T =x$. Thus, in either case, $y \ge x$, contradicting $y < x-1$. Similarly, if $b_j$ is unmatched, Properties~\ref{property:GM}(\ref{obs:partitionPorp5}) and~\ref{property:GM}(\ref{obs:partitionPorp6}) imply $x \le y$. Thus, both clones must be matched.

Let $z = x-1$. Since $a_i\in\AA_x$ is matched, $a$ reached level $x$ and hence exhausted its proposal list at level $z$. If $z < \T$, the matched clone $b_j \in \BB_y$, where $y < z < \T$, shows that $b$ is an \emph{lq}-vertex and hence belongs to $\prefslqa$. If $z \ge \T$, the proposal list contains all neighbors of $a$. Therefore, $a$ proposed to $b$ at level $z$.

Since $(a,b)\notin M$, this proposal was rejected immediately or subsequently. Therefore, all partners of $b$ have levels at least $z$. Moreover, $b$ has at least $q^-(b)$ partners if $z < \T$, and is fully subscribed relative to $q^+(b)$ if $z\ge\T$. Here, a starred replacement can occur only after its proposer's earlier normal proposal was rejected, immediately or subsequently, so Observation~\ref{obs:algorithm-invariants} (\ref{obs:normal-rejection}) applies. All these conditions continue to hold thereafter.
Thus, $b$ cannot have a final partner at level $y < z$, contradicting $b_j \in \BB_y$.
\end{proof}

\begin{lemma}\label{lem:noSteepMM}
The graph $G_M$ contains no steep downward edge. That is, there is no edge between $\AA_x$ and $\BB_y$ with $x>y+1$.
\end{lemma}

\begin{proof}
Every edge of $M^*$ is horizontal. Every edge of $E'_U$ represents a pair $(a,b)\in E\setminus M$. If its endpoint in $\AA'$ belongs to $\AA_x$ with $x\ge1$, Claim~\ref{cl:noedgeMM} implies that its other endpoint is in $\BB_y$ with $y \ge x-1$. For $x=0$, this inequality holds automatically. Hence, no edge is steep downward.
\end{proof}

We next bound the number of partitions containing clones of a vertex that never reaches a level above $\T$.

\begin{cl}\label{cl:twoConscutiveL}
Let $h\le\T$ be the highest level reached by $a \in \AA$, treating sub-level $\T^*$ as level $\T$. Then all clones of $a$ belong to $\AA_{\max \{0, h-1\}} \cup \AA_h$. Thus, they are in at most two consecutive partitions.
\end{cl}

\begin{proof}
If $h \ge 1$, then before reaching level $h$, $a$ exhausts its proposal list at level $h-1$. Since its proposal lists at lower levels are already exhausted, no edge with a label below $h-1$ is subsequently introduced. Hence, every matched clone of $a$ belongs to $\AA_{h-1} \cup \AA_h$. If $h = 0$, every matched clone belongs to $\AA_0$.

If $h<\T$, then $a$ is fully subscribed at termination; otherwise, it would continue proposing and eventually advance beyond $h$. If $h = \T$, then $a$ is non-deficient at termination; otherwise, it would advance to level $\T + 1$. Thus, any unmatched clones are non-critical and belong to $\AA_{\T} = \AA_h$. This proves the claim.
\end{proof}

We next relate blocking pairs in $G$ to those in $G_M$. Each clone inherits the original preference ordering of its vertex. That is, among its neighbors in $G_M$, clones of the same original vertex are tied, and comparisons between clones of different vertices follow the original preference list.

\begin{lemma}\label{cl:blockingUpwards}
Let $(a,b)$ be a blocking pair with respect to $M$ in $G$. Then a blocking pair $(a_i,b_j)$ exists with respect to $M^*$ in $G_M$. Moreover, every blocking pair $(a_i,b_j)$ with respect to $M^*$ is upward; that is, if $a_i \in \AA_x$ and  $b_j \in \BB_y$, then $x < y$.
\end{lemma}

\begin{proof}
Let $(a,b)$ block $M$. If $a$ is under-subscribed, choose an unmatched clone $a_i$; otherwise, choose the clone matched to a representative of some $b' \in M(a)$ with $b \succ_a b'$. Choose $b_j$ analogously. Since $(a,b) \in E \setminus M$, the edge $(a_i,b_j)$ belongs to $G_M$. Each endpoint is either unmatched or strictly prefers the other endpoint to its partner. Thus, $(a_i, b_j)$ blocks $M^*$.

Consider a blocking pair $(a_i,b_j)$ with $a_i \in \AA_x$ and $b_j \in \BB_y$, and suppose $y \le x$. If $x < \T$, then $y \le x <\T$ implies that $b$ is an \emph{lq}-vertex; otherwise, the proposal list contains all neighbors of $a$. Clearly, the blocking condition implies that $a$ proposed to $b$ at level $x$. Since $(a,b) \notin M$, the proposal was rejected immediately or eventually. We show that, from this rejection onward, all partners of $b$ have levels at least $x$, and those at level $x$ are weakly preferred to $a$.

If $x = \T$, then $a$ has passed the rank of $b$ or exhausted level $\T$. Its last level-$\T$ proposal cannot be rejected by the uncertain-proposal rule because, before passing this rank, such a rejection forces a marked reproposal, and afterward, its uncertain label has been cleared. The rejection rules and Observation~\ref{obs:uncertainty-invariants} therefore ensure that $b$ is fully subscribed, does not hold any uncertain proposals at the time of that rejection, and remains so.

If $x \ne \T$, the rejection rules 
ensure that all partners of $b$ have levels at least $x$, and those at level $x$ are weakly preferred to $a$, unless the rejection is uncertainty-based. For $x < \T$, such a rejection requires that $a$ first propose again to $b$ at level $\T$ and upgrade the pair. At that rejection, all partners of $b$ have levels at least $\T > x$, since partners below $\T$ are rejected first. For $x > \T$, an uncertainty-based rejection is impossible. These two properties are preserved by subsequent proposals.

At the time when $b$ rejected $a$, it had at least $q^-(b)$ partners if $x < \T$, and was fully subscribed otherwise. Therefore, if $x < \T$, all critical clones of $b$ are matched, and any unmatched clone belongs to $\BB_\T$; therefore, $b_j$ must be matched because $y \le x < \T$. If $x \ge \T$, all clones of $b$ are matched.

Since all matched clones of $b$ belong to partitions with indices at least $x$, we have $y \ge x$. Together with the assumption $y \le x$, we get $y = x$. Every partner of $b$ at level $x$ is weakly preferred to $a$. By the inherited preferences, $b_j$ therefore weakly prefers its $M^*$ partner to $a_i$, contradicting that $(a_i,b_j)$ is a blocking pair. Therefore, $x<y$.
\end{proof}

\subsubsection{Criticality and relaxed stability of our matching}

We now show that the output matching $M$ is critical and relaxed-stable. For this, we define some terminology that will be helpful. Given a matching $M$, let $\Dfa{M}$ and $\Dfb{M}$ denote the total deficiencies of the vertices in $\AA$ and $\BB$, respectively. That is,
\[
\Dfa{M} = \sum_{a \in \AA} \max\{0, q^-(a)-|M(a)|\},
\qquad
\Dfb{M} = \sum_{b \in \BB} \max\{0, q^-(b)-|M(b)|\}.
\]
Equivalently, these quantities count the unfilled critical positions on the respective sides. The total deficiency of $M$ is therefore
\[
\Df{M}=\Dfa{M}+\Dfb{M}.
\]

\begin{lemma}\label{lem:CriticalM}
The output matching $M$ of our algorithm is critical for $G$.
\end{lemma}

\begin{proof} 
 Let $N$ be any critical matching in $G$. Since $N$ is critical, $\Df{N}$ is the minimum deficiency among all possible matchings. We show that there is no alternating path $\rho$ in $G_M$ with respect to $M^*$ such that $M^*\oplus \rho$ has a lower deficiency than $M$. To this end, we map the many-to-many matching $N$ to a corresponding one-to-one matching $N^*$ in the cloned graph $G_M$. First, we observe that the cloned graph $G_M$ can be constructed in such a way that the number of unmatched critical clones in $N^*$ is equal to $\Df{N}$ and the number of unmatched critical clones in $M^*$ is equal to $\Df{M}$. This can be done as follows. We assume $M^*=N^*=\emptyset$ initially in the following construction. 
 \begin{itemize}
    \item[-]  For every edge $(a, b) \in E \setminus M$, we have all edges corresponding to the complete bipartite graph between the clones of $a$, and clones of $b$. 
    \item[-] For every edge $(a, b)\in M \cap N$ we select a lowest index  clone of $a$, say $a_i$ unmatched in $M^*$ as well as $N^*$, and a lowest index clone of $b$, say $b_j$ that is unmatched in both $M^*$ and $N^*$. Set $M^*=M^*\cup\{(a_i, b_j)\}$,  and $N^*=M^*\cup\{(a_i, b_j)\}$. Note that the lowest indexed clones are used for the edges in the matching $M\cap N$.
    \item[-] For every edge $(a, b)\in M\setminus N$ we select a lowest index  clone of $a$, say $a_i$, and a lowest index clone of $b$, say $b_j$ where both clones are unmatched in $M^*$. Set  $M^*=M^*\cup\{(a_i, b_j)\}$.
     
     \item[-] For every edge $(a,b)\in N\setminus M$, we find the lowest index clone $a_i$ of $a$ and the lowest index clone $b_j$ of $b$ where both clones are  unmatched in $N^*$. Set  $N^*=N^*\cup\{(a_i, b_j)\}$.
 \end{itemize}
 The partition of the clones in $G_M$ is exactly as described in Section~\ref{sec:cloneG}.
 
 Now, we proceed to prove the criticality of matching $M$. The proof is divided into two parts called the $\AA$-part and the $\BB$-part. In the $\AA$-part, we show that $\Dfa{M}\le \Dfa{N} $, and in the $\BB$-part, we show that $\Dfb{M}\le \Dfb{N}$. These together imply that $M$ is a critical matching.
 
 \vspace{0.1in}
 
 \noindent \textbf{Proof of ($\AA$-part):} For the sake of contradiction, let us assume that $M$ is not critical for the $\AA$-part, that is, $\Dfa{N}<\Dfa{M}$. This implies that there exists a maximal alternating path $\rho$ in $M^*\oplus N^*$ such that $N^*$ matches more critical clones in $\rho\cap\AA'$ than the matching $M^*$. 
 The idea is to show that 
 the number of critical clones in $\rho \cap \AA'$ is more than the number of critical clones in $\AA'$ which is a contradiction. 
Consider the case when such an alternating path $\rho$ ends in $\AA'$. 
Let $\rho=\langle u_0, v_1,u_1,v_2,u_2,\ldots,v_k,u_k\rangle$,
where $(v_i,u_i)\in M^*$ and the other edges of $\rho$ are in the matching $N^*$. Here, $u_0$ is a critical clone of some vertex $a\in\AA$, and $u_0$ is unmatched in $M^*$. 
Also, by our choice of $\rho$, $N^*$ matches more critical clones than $M^*$ on $\rho$. Hence $u_k$ is a non-critical clone of some $a' \in \AA$ where $a' \neq a$. By the level-wise partition of the clones, $u_0 \in \AA_{\S+\T}$ and $u_k \in \AA_{x}$ for $x\le \T$. 

Since $u_0\in\AA_{\S+\T}$ and $u_0$ is unmatched in $M^*$ , Property~\ref{property:GM}(\ref{obs:partitionPorp3}) implies that all neighbors of $u_0$ are only in $\BB_{\S+\T}$. Hence,  $v_1\in \BB_{\S+\T}$. Since $(u_1,v_1)\in M$, $u_1$ must be in $\AA_{\S+\T}$ (see Figure~\ref{fig:critLem}). 

\begin{figure}
\begin{center}
    \scalebox{0.85}{\begin{tikzpicture}[thick,
  lnode/.style={draw=white,fill=blue},
  fsnode/.style={draw,circle,fill=black,scale=0.5},
  every fit/.style={ellipse,draw,inner sep=-2pt,text width=0.5cm},
  -,shorten >= 3pt,shorten <= 3pt
]


  \node[fsnode] (u0) at (0,0.5) {};
  \node at (-0.5,0.5) {$u_0$};
  \node[fsnode] (v1) at (2,0) {};
  \node at (2.5,0) {$v_1$};
  \node[fsnode] (u1) at (0,0) {};
  \node at (-0.5,0) {$u_1$};
  \node at (-2,0.3) {$\AA_{\S+\T}$};
  \node at (4,0.3) {$\BB_{\S+\T}$};
   \draw[ultra thin, dashed](-1.5,-0.5) -- (3.5,-0.5);
   
  \node[fsnode] (v2) at (2,-1) {};
  \node at (2.5,-1) {$v_2$};
  \node[fsnode] (u2) at (0,-1) {};
  \node at (-0.5,-1) {$u_2$};
   \node at (-2,-1) {$\AA_{\S+\T-1}$};
   \node at (4,-1) {$\BB_{\S+\T-1}$};
   \draw[ultra thin,dashed](-1.5,-1.5) -- (3.5,-1.5);
   
   \node[fsnode] (ui1) at (0,-1.75) {};
   \node[fsnode] (vi1) at (2,-1.75) {};
   \node at (1,-1.75) {$\ldots$};
   \draw[ultra thin,dashed](-1.5,-2) -- (3.5,-2);
   
  \node[fsnode] (vi) at (2,-2.5) {};
  \node at (2.5,-2.5) {$v_i$};
  \node[fsnode] (ui) at (0,-2.5) {};
  \node at (-0.5,-2.5) {$u_i$};
   \node at (-2,-2.5) {$\AA_{j}$};
  \node at (4,-2.5) {$\BB_{j}$};
   \draw[ultra thin, dashed](-1.5,-3) -- (3.5,-3);
   
   \node[fsnode] (ui11) at (0,-3.25) {};
   \node[fsnode] (vi11) at (2,-3.25) {};
   \node at (1,-3.25) {$\ldots$};
   \draw[ultra thin, dashed](-1.5,-3.5) -- (3.5,-3.5);

\node[fsnode] (vk1) at (2,-4) {};
  \node[fsnode] (uk1) at (0,-4) {};
   \node at (-2,-4) {$\AA_{\T+1}$};
  \node at (4,-4) {$\BB_{\T+1}$};
   \draw[ultra thin, dashed](-1.5,-4.5) -- (3.5,-4.5);
   
   \node[fsnode] (vk) at (2,-5) {};
  \node[fsnode] (uk) at (0,-5) {};
   \node at (-2,-5) {$\AA_{\le \T}$};
  \node at (4,-5) {$\BB_{\le \T}$};
  
  \node[fsnode] (vk11) at (2,-5.5) {};
  \node at (2.5,-5.5) {$v_{k}$};
  \node[fsnode] (uk11) at (0,-5.5) {};
  \node at (-0.5,-5.5) {$u_{k}$};

 \draw[thick, red ](u0) -- (v1);
 \draw[thick, blue](u1) -- (v1);
 \draw[thick, red ](u1) -- (v2);
 \draw[thick, blue](u2) -- (v2);
 \draw[thick, blue](ui) -- (vi);
 \draw[thick, red ](u2) -- (vi1);
 \draw[thick, red ](ui1) -- (vi);
 \draw[thick, red ](ui) -- (vi11);
 \draw[thick, red ](ui11) -- (vk1);
 \draw[thick, blue](uk1) -- (vk1);
 \draw[thick, red ](uk1) -- (vk);
 \draw[thick, dotted,blue](uk) -- (vk);
 \draw[thick, dotted, red ](uk) -- (vk11);
 \draw[thick, dotted,blue](uk11) -- (vk11);

\end{tikzpicture}}
\end{center}

\caption{Blue colored edges denote the edges in $M^*$ whereas red edges denote the edges in $N^*$. }
\label{fig:critLem}
\end{figure}


By Lemma~\ref{lem:noSteepMM}, we know that $G_M$ does not contain steep downward edges. Hence, if $u_i\in\AA_x$ and $u_{i+1}\in\AA_y$ then $x-y\le 1$ for all indices $i$ on $\rho$ and hence, $\rho$ must contain at least one clone from each $\AA_x$ for $\T+1\le x\le \S+\T$ (see Figure~\ref{fig:critLem}). Now notice that there are at least two clones $u_0$ and $u_1$ in $\AA_{\S+\T}$ and at least one clone in each $\AA_x$ for $\T+1\le x\le \S+\T-1$. As all clones in $\bigcup_{x=\T+1}^{\S+\T}\AA_x$ are critical, there are at least $\S+1$ many critical clones in $\rho\cap \AA'$ contradicting the fact that there are  $s$ critical clones in $\AA'$.

The other case is when $\rho$ ends in $\BB'$, say at $v_{k+1}$. Then $v_{k+1}$ is unmatched in $M^*$. If it is critical, it belongs to $\BB_0$, and Property~\ref{property:GM}(\ref{obs:partitionPorp5}) gives $u_k\in\AA_0$. Otherwise, it belongs to $\BB_{\T}$, and Property~\ref{property:GM}(\ref{obs:partitionPorp6}) gives $u_k \in \bigcup_{x=0}^{\T} \AA_x$. Thus, in either case, $u_k$ lies at or below $\T$, and the same counting argument as above yields a contradiction.

 \vspace{0.1in}

\noindent \textbf{Proof of ($\BB$-part):}
Proof of this part is similar to that of $\AA$-part. For the sake of contradiction, let us assume that $\Dfb{N}<\Dfb{M}$. This implies that there exists an alternating path $\rho$ in $M^*\oplus N^*$ such that $N^*$ matches more critical clones in $\rho\cap\BB'$ than the matching $M^*$. It can be shown that the number of critical clones in $\rho \cap \BB'$ is more than the number of critical clones in $\BB'$, which is a contradiction. 

 The path $\rho$ must start at a clone in $\BB_0$ must end at a clone in $\BB_x $ where $x \ge t$ or at a clone in $\AA_t \cup \AA_{s+t}$. As above, in either case, the path has at least one critical clone in each of  $\BB_1, \ldots, \BB_{t-1}$ and at least two clones in $\BB_0$ accounting for a total of $t+1$ critical clones in $\rho \cap \BB'$, a contradiction to the number of critical clones in $\BB'$.
\end{proof}

\begin{lemma}\label{lem:rwsmMM}
The output matching $M$ is \RSM\ for $G$.
\end{lemma}

\begin{proof}
If there is no blocking pair with respect to $M$, there is nothing to prove. So, let $(a,b)$ be a blocking pair with respect to $M$. It suffices to consider the case where it is not justified from the $a$-side. In this case, we identify a clone $a_i \in \AA_x$ with $x \ge \T$ such that $a_i$ is unmatched or strictly prefers the clones of $b$ to its partner.  Using this clone, we show that $b$ is fully subscribed and that every partner of $b$ strictly worse than $a$ is non-surplus, thereby justifying $(a,b)$ from the $b$-side.

If $a$ is under-subscribed, choose an unmatched clone $a_i$. By construction, it belongs to $\AA_\T$ or $\AA_{\S+\T}$, so $x \ge \T$. 
Otherwise, $a$ is fully subscribed. If every $b' \in M(a)$ with $b \succ_a b'$ were non-surplus, then the blocking pair $(a,b)$ would be justified from the $a$-side, contrary to our assumption. Hence, there exists $b'\in M(a)$ such that $b \succ_a b'$ and $|M(b')| > q^-(b')$. Choose the clone $a_i$ matched to a clone of $b'$. Since $b \succ_a b'$, this clone strictly prefers every clone of $b$ to its partner. Moreover, Observation~\ref{obs:algorithm-invariants}(\ref{obs:temporary-capacity}) implies that $b'$ cannot have a partner below level $\T$, since it is surplus. Thus, the edge representing $(a,b')$ is at level at least $\T$, giving $x \ge \T$.

Since $(a,b) \notin M$, $a_i$ is adjacent to every clone of $b$. If $b$ were under-subscribed, an unmatched clone $b_j$ would belong to $\BB_0$ or $\BB_\T$. Then $(a_i,b_j)$ would block $M^*$ without being upward, contradicting Lemma~\ref{cl:blockingUpwards}. Hence, $b$ is fully subscribed.

It remains to show that every partner of $b$ that is strictly worse than $a$ is non-surplus. Let $a' \in M(b)$ satisfy $a \succ_b a'$, and let $(a'_r,b_j)\in M^*$ represent the edge $(a',b) \in M$. By our choice, $a_i$ is unmatched or strictly prefers $b_j$ to its partner. Also, since $a \succ_b a'$, the clone $b_j$ strictly prefers $a_i$ to its partner $a'_r$. Thus, $(a_i,b_j)$ blocks $M^*$. Writing $b_j \in \BB_y$, Lemma~\ref{cl:blockingUpwards} gives $y > x \ge \T$.  Thus, $(a',b)$ is at level $y > \T$, so $a'$ reached a level above $\T$. By the bound established in the proof of Property~\ref{property:GM}(\ref{obs:partitionPorp1}), $|M(a')|\le q^-(a')$.

Since $a'$ was arbitrary, every partner of $b$ strictly worse than $a$ is non-surplus. Together with the fact that $b$ is fully subscribed, this justifies the blocking pair $(a, b)$ from the $b$-side. Thus, every blocking pair is justified from at
least one side, and $M$ is \RSM.
\end{proof}

\begin{remark}\label{remark:Csaji}
Cs{\'a}ji~\cite{csaji2024AAMAS} introduced a stronger notion of critical relaxed stability in the one-to-one setting. Under this notion, a matching $M$ must be critical and admit no blocking pair $(a,b)$ for which matching $a$ to $b$ and rejecting their existing partners, if any, preserves criticality. He proved that a critical relaxed-stable matching always exists under this stronger definition. Lemma~\ref{lem:equivCsaji} below shows that the output of our algorithm satisfies the corresponding strengthening in the many-to-many setting.
\end{remark}

\begin{lemma}\label{lem:equivCsaji}
Let $M$ be the output of Algorithm~\ref{algo:maxCRSMM}. There is no edge $(a,b) \in E \setminus M$ with $a'\in M(b) \cup \{\bot\}$ and $b'\in M(a) \cup \{\bot\}$ satisfying $a\succ_b a'$ and $b\succ_a b'$ for which the matching $N = (M \setminus \{(a,b'), (a',b)\}) \cup \{(a,b)\}$ is critical. Here, $\bot$ may be chosen only for an under-subscribed endpoint. Every vertex strictly prefers each acceptable neighbor to $\bot$, and pairs involving $\bot$ are omitted when defining a matching.
\end{lemma}

\begin{proof}
Suppose, for a contradiction, that such choices $(a,b)$, $a'$, and $b'$ exist and $N = (M \setminus \{(a,b'), (a',b)\}) \cup \{(a,b)\}$ is critical. The conditions of the lemma imply that $(a,b)$ blocks $M$. Since both $M$ and $N$ are critical, $\Df{N} = \Df{M}$. We consider the following cases.

\begin{enumerate}[left=10pt, label=Case \arabic*.]
    \item \label{itm:case1} \textbf{$a'$ is either $\bot$ or surplus in $M$, and $b'$ is either $\bot$ or surplus in $M$.} If $b' = \bot$, then $a$ is under-subscribed. Otherwise, $a$ strictly prefers $b$ to its surplus partner $b'$. In either case, $(a,b)$ is not justified from the $a$-side. Similarly, it is not justified from the $b$-side, contradicting Lemma~\ref{lem:rwsmMM}.

    \item \label{itm:case2} \textbf{$b' \ne \bot$ and $b'$ is non-surplus in $M$.} In $N$, $b'$ loses one partner, increasing its deficiency by one. The number of partners of $a$ is unchanged, and only $b$ can have a smaller deficiency. Thus, $\Df{N}=\Df{M}$ implies that $a' = \bot$ and $b$ is deficient in $M$.

    Let $a_i$ be the clone of $a$ matched to a clone of $b'$, and choose an unmatched critical clone $b_j \in \BB_0$ of $b$. Since $(a,b) \notin M$, Property~\ref{property:GM}(\ref{obs:partitionPorp5}) implies $a_i \in \AA_0$. Moreover, $b \succ_a b'$, so $(a_i, b_j)$ blocks $M^*$. This horizontal blocking pair contradicts Lemma~\ref{cl:blockingUpwards}.

    \item \label{itm:case3} \textbf{$a'\ne\bot$ and $a'$ is non-surplus in $M$.} Here, $a'$ loses one partner, increasing its deficiency by one. Since the number of partners of $b$ is unchanged, only $a$ can have a smaller deficiency. Hence, $\Df{N} = \Df{M}$ implies that $b' = \bot$ and $a$ is deficient in $M$.

    Choose an unmatched critical clone $a_i \in \AA_{\S+\T}$ of $a$, and let $b_j$ be the clone of $b$ matched to a clone of $a'$. Property~\ref{property:GM}(\ref{obs:partitionPorp3}) implies $b_j \in \BB_{\S + \T}$. Since $a\succ_b a'$, the pair $(a_i, b_j)$ blocks $M^*$. Again, this horizontal blocking pair contradicts Lemma~\ref{cl:blockingUpwards}.
\end{enumerate}
These cases exhaust all possibilities, proving the lemma.
\end{proof}

\begin{lemma}\label{lem:2by3RSMmm}
Let $M$ be the output of Algorithm~\ref{algo:maxCRSMM}, and let $M_{max}$ be a maximum-size \RSM\ in $G$. Then $|M|\ge \frac{2}{3}|M_{max}|$.
\end{lemma}

\begin{proof}
Construct the corresponding one-to-one matchings $M^*$ and $M_{max}^*$ in $G_M$ as in the proof of Lemma~\ref{lem:CriticalM}, processing common edges first and using the lowest-index available clones. This construction does not require $M_{max}$ to be critical. Common edges have the same representatives, and each matching fills initial clones of every vertex.

We show that $M^* \oplus M_{max}^*$ contains no augmenting path of length one or three with respect to $M^*$. Since $M$ is \RSM, it is maximal. Thus, no edge outside $M$ joins two vertices that are under-subscribed in $M$, ruling out an augmenting path of length one.

Suppose, for a contradiction, that there is an
augmenting path $\rho = \langle \widetilde{a}_1, b_1, a_1, \widetilde{b}_1 \rangle$, where $(a_1, b_1) \in M^*$ and the other two edges belong to $M_{max}^*$. Let these clones correspond to $\widetilde{a}, b,a,\widetilde{b}$, respectively. By the alignment of common edges, $(a,b) \in M \setminus M_{max}$ and $(\widetilde{a}, b),(a, \widetilde{b}) \in M_{max} \setminus M$. Both $\widetilde{a}$ and $\widetilde{b}$ are under-subscribed in $M$.

\medskip
\noindent\textbf{The highest level reached by $a$ is $\T$.} Since $\widetilde{a}$ is under-subscribed, Property~\ref{property:GM}(\ref{obs:partitionPorp4}) implies that $b_1$ belongs to a partition with index at least $\T$. Similarly, since $\widetilde{b}$ is under-subscribed and $(a, \widetilde{b}) \notin M$, Property~\ref{property:GM}(\ref{obs:partitionPorp6}) implies that $a_1$ belongs to a partition with index at most $\T$. Since $(a_1, b_1) \in M^*$, its endpoints have the same partition index. Hence, $a_1 \in \AA_\T$ and $b_1 \in \BB_\T$.

Moreover, $a$ never proposed to $\widetilde{b}$ at level $\T$. Otherwise, since $\widetilde{b}$ remains under-subscribed, that proposal would have been accepted and the pair could not subsequently have been rejected, by Observation~\ref{obs:algorithm-invariants}(\ref{obs:normal-rejection}). Thus, $a$ never entered sub-level $\T^*$, so the highest level reached by $a$ is $\T$. Since $a^\T$ proposed to $b$ but not to $\widetilde{b}$, the favorite-neighbor rule implies $b \succeq_a \widetilde{b}$.

\medskip
\noindent\textbf{The pair $(a,b)$ blocks $M_{max}$.} Since $\widetilde{a}$ is under-subscribed in $M$, it exhausted sub-level $\T^*$. We first show that $a \succ_b \widetilde{a}$ and that $b$ holds no uncertain proposal in the final matching.

Consider the proposal from $\widetilde{a}^{\T^*}$ to $b$. If it was rejected immediately, then $\widetilde{a}$'s last level-$\T$ proposal to $b$ had already been rejected. This earlier rejection was not uncertainty-based because such a rejection would force a marked reproposal before leaving level $\T$, whereas uncertain labels are cleared upon leaving. The rejection rules and Observation~\ref{obs:uncertainty-invariants} therefore imply that $b$ was fully subscribed, held no uncertain proposal, and weakly preferred every partner at level $\T$ to $\widetilde{a}$. These conditions continue to hold thereafter. Consequently, the starred proposal could be rejected only if every unstarred partner at level $\T$ was strictly preferred to $\widetilde{a}$. 

If the proposal from $\widetilde{a}^{\T^*}$ to $b$ was accepted, then it was subsequently rejected because $(\widetilde{a}, b) \notin M$. At that rejection, $b$ was fully subscribed and held no uncertain proposal. Indeed, a level-$\T$ proposal rejects an uncertain partner first, and Observation~\ref{obs:uncertainty-invariants}(\ref{obs:advanced-proposals}) excludes a higher-level proposal from a non-partner while an uncertain proposal is held. The comparisons by level, preference rank, and starred status then ensure that every remaining unstarred partner at level $\T$ is strictly preferred to $\widetilde{a}$.

In either case, $b$ remains fully subscribed, receives no new uncertain labels, and preserves this strict preference for its unstarred partners at level $\T$. Since $(a,b)\in M$ and $a$ never entered sub-level $\T^*$, we obtain $a \succ_b \widetilde{a}$.

It remains to show that $b \succ_a \widetilde{b}$. Suppose that $b$ and $\widetilde{b}$ are tied. The level-$\T$ proposal establishing the final pair $(a, b)$ must be a first proposal, since the unproposed, under-subscribed neighbor $\widetilde{b}$ would be selected before a marked $b$. The uncertainty test therefore labels $(a^\T, b)$ as uncertain. Since $a$ never proposes to $\widetilde{b}$  at level $\T$, it neither passes this rank nor enters sub-level $\T^*$, so this uncertain label holds until termination. This contradicts that $b$ holds no uncertain proposal in the final matching. 

Thus, $b \succ_a \widetilde{b}$ and $a \succ_b \widetilde{a}$. Since $(a, \widetilde{b}), (\widetilde{a},b) \in M_{max}$, the pair $(a,b)$ blocks $M_{max}$.

\medskip
\noindent\textbf{This blocking pair $(a,b)$ for $M_{max}$ is not justified.} The matching $P = (M \setminus \{(a, b)\}) \cup \{(\widetilde{a}, b), (a, \widetilde{b})\}$ respects the upper quotas because $\widetilde{a}$ and $\widetilde{b}$ are under-subscribed in $M$. The number of partners of each of $a$ and $b$ is unchanged, whereas each of $\widetilde{a}$ and $\widetilde{b}$ gains one partner. All other vertices are unaffected.

If $|M(\widetilde{a})| < q^-(\widetilde{a})$, then the deficiency of $\widetilde{a}$ decreases by one, while no vertex's deficiency increases. This gives $\Df{P} < \Df{M}$, contradicting the criticality of $M$. The same argument applies to $\widetilde{b}$. Therefore, we have $|M(\widetilde{a})| \ge q^-(\widetilde{a})$ and $|M(\widetilde{b})| \ge q^-(\widetilde{b})$. Each endpoint clone of $\rho$ is matched in $M_{max}^*$ but unmatched in $M^*$. Since both matchings fill initial clones of each vertex, we get 
\[
|M_{max}(\widetilde{a})| > |M(\widetilde{a})| \ge q^-(\widetilde{a}) \qquad \text{and } \qquad
|M_{max}(\widetilde{b})| > |M(\widetilde{b})| \ge q^-(\widetilde{b}).
\]
Hence, both $\widetilde{a}$ and $\widetilde{b}$
are surplus in $M_{max}$. The blocking pair $(a,b)$ is therefore not justified from either side, contradicting that $M_{max}$ is \RSM.

Thus, every augmenting path in $M^* \oplus  M_{max}^*$ has length at least five. Such a path contains $k$ edges of $M^*$ and $k+1$ edges of $M_{max}^*$ for some $k \ge 2$, so $k+1 \le \frac{3}{2} k$. Every other component contains at most as many edges of $M_{max}^*$ as of $M^*$. Therefore, $|M_{max}| = |M_{max}^*| \le \frac{3}{2} |M^*| = \frac{3}{2}|M|$.
\end{proof}

Lemmas~\ref{lem:CriticalM}, \ref{lem:rwsmMM}, and~\ref{lem:2by3RSMmm} together prove Theorem~\ref{theo:main}.

\section{Conclusion}


In this paper, we study the many-to-many bipartite matching problem with two-sided preferences, where preferences may have ties and lower quotas are present for agents on both sides of the bipartition. We prove that a critical relaxed-stable matching always exists and can be computed in polynomial time. Finding a maximum-size such matching is NP-hard even when all lower quotas are zero, in which case critical relaxed stability coincides with weak stability. We present a polynomial-time algorithm that computes a critical relaxed-stable matching whose size is at least $\frac{2}{3}$ times the size of a maximum-size relaxed-stable matching. Thus, the output minimizes total deficiency, while its size guarantee is measured against \textit{all} relaxed-stable matchings, including those that are not critical. Our approximation factor matches the best-known bound for maximum-size weakly stable matching in the presence of ties~\cite{mcdermid20093,Kiraly13,paluch2014faster}.

We further show that the output satisfies the many-to-many extension of the stronger notion of critical relaxed stability introduced by Cs{\'a}ji~\cite{csaji2024AAMAS}. Hence, both the existence result and the size guarantee hold under this strengthened notion. We also establish that deciding whether a stable critical matching exists is NP-complete, even in the one-to-one setting with complete preference lists and lower quotas on only one side of the bipartition. In addition, we correct a technical issue in the definition of uncertain proposals in Kir{\'a}ly's algorithm~\cite{Kiraly13} and extend the corrected framework to the many-to-many setting.

 \bibliographystyle{alphaurl} 
 \bibliography{refs.bib}

@article{cseh2017popular,
  title={Popular matchings},
  author={Cseh, {\'A}gnes},
  journal={Trends in Computational Social Choice},
  volume={105},
  number={3},
  year={2017},
  publisher={Lulu. com}
}

@article{Simple15Approximation,
  author    = {Gergely Cs{\'a}ji and Tam{\'a}s Kir{\'a}ly and Yu Yokoi},
  title     = {A Simple 1.5-Approximation Algorithm for a Wide Range of Maximum-Size Stable Matching Problems},
  journal   = {Mathematics of Operations Research},
  year      = {2025},
  doi       = {10.1287/moor.2024.0725},
  url       = {https://doi.org/10.1287/moor.2024.0725}
}

@article{CEFMMP2014,
  title={Pareto optimality in many-to-many matching problems},
  author={Cechl{\'a}rov{\'a}, Katar{\'\i}na and Eirinakis, Pavlos and Fleiner, Tam{\'a}s and Magos, Dimitrios and Mourtos, Ioannis and Potpinkov{\'a}, Eva},
  journal={Discrete Optimization},
  volume={14},
  pages={160--169},
  year={2014},
  publisher={Elsevier},
url          = {https://doi.org/10.1016/j.disopt.2014.09.002}
}

@inproceedings{dudycz2022tight,
  title={Tight {I}napproximability of {M}inimum {M}aximal {M}atching on {B}ipartite {G}raphs and {R}elated {P}roblems},
  author={Dudycz, Szymon and Manurangsi, Pasin and Marcinkowski, Jan},
  booktitle={Approximation and Online Algorithms: 19th International Workshop, WAOA 2021, Lisbon, Portugal, Revised Selected Papers},
  pages={48--64},
  year={2022},
  organization={Springer},
url          = {https://doi.org/10.1007/978-3-030-92702-8\_4}
}

@inproceedings{MakinoMY22,
  author       = {Kazuhisa Makino and
                  Shuichi Miyazaki and
                  Yu Yokoi},
  title        = {Incomplete {L}ist {S}etting of the {H}ospitals/{R}esidents {P}roblem with {M}aximally
                  {S}atisfying {L}ower {Q}uotas},
  booktitle    = {15th International Symposium on Algorithmic Game Theory, {SAGT} 2022,              Colchester, UK},
  pages        = {544--561},
  publisher    = {Springer},
  year         = {2022},
  url          = {https://doi.org/10.1007/978-3-031-15714-1\_31}
}

@article{irving1994stable,
  author       = {Robert W. Irving},
  title        = {Stable Marriage and Indifference},
  journal      = {Discret. Appl. Math.},
  volume       = {48},
  number       = {3},
  pages        = {261--272},
  year         = {1994},
  url          = {https://doi.org/10.1016/0166-218X(92)00179-P},
  doi          = {10.1016/0166-218X(92)00179-P}
}

@inproceedings{RaghavendraSmallSet,
author = {Raghavendra, Prasad and Steurer, David},
title = {{Graph Expansion and the Unique Games Conjecture}},
year = {2010},
url = {https://doi.org/10.1145/1806689.1806792},
booktitle = {Proceedings of the Forty-Second ACM Symposium on Theory of Computing, (STOC 2010)},
pages = {755–764},
Address = {Cambridge, Massachusetts, USA}
}

@article{manlove2002hard,
  title={Hard variants of stable marriage},
  author={Manlove, David F and Irving, Robert W and Iwama, Kazuo and Miyazaki, Shuichi and Morita, Yasufumi},
  journal={Theoretical Computer Science},
  volume={276},
  number={1-2},
  pages={261--279},
  year={2002},
  publisher={Elsevier},
url          = {https://doi.org/10.1016/S0304-3975(01)00206-7}
}

@inproceedings{GokoMMY22,
  author       = {Hiromichi Goko and
                  Kazuhisa Makino and
                  Shuichi Miyazaki and
                  Yu Yokoi},
  title        = {Maximally {S}atisfying {L}ower {Q}uotas in the {H}ospitals/{R}esidents {P}roblem
                  with {T}ies},
  booktitle    = {39th International Symposium on Theoretical Aspects of Computer Science, {STACS} 2022, Marseille, France (Virtual Conference)},
  pages        = {31:1--31:20},
  year         = {2022},
  url          = {https://doi.org/10.4230/LIPIcs.STACS.2022.31}
}

@inproceedings{mcdermid20093,
  title={A 3/2-{A}pproximation {A}lgorithm for {G}eneral {S}table {M}arriage},
  author={McDermid, Eric},
  booktitle={International Colloquium on Automata, Languages, and Programming},
  pages={689--700},
  year={2009},
  organization={Springer},
url          = {https://doi.org/10.1007/978-3-642-02927-1\_57}
}

@inproceedings{NN17,
  author    = {Meghana Nasre and
               Prajakta Nimbhorkar},
  title     = {Popular {M}atchings with {L}ower {Q}uotas},
  booktitle = {37th {IARCS} Annual Conference on Foundations of Software Technology
               and Theoretical Computer Science (FSTTCS 2017)},
  pages     = {44:1--44:15},
  year      = {2017},
  url       = {https://doi.org/10.4230/LIPIcs.FSTTCS.2017.44},
}

@inproceedings{Kavitha2021,
  author    = {Telikepalli Kavitha},
  title     = {Matchings, {C}ritical {N}odes, and {P}opular {S}olutions},
  booktitle = {41st {IARCS} Annual Conference on Foundations of Software Technology
               and Theoretical Computer Science (FSTTCS 2021)},
  pages     = {25:1--25:19},
  year      = {2021},
  url          = {https://doi.org/10.4230/LIPIcs.FSTTCS.2021.25}
}

@article{krishnaa2023envy,
  title={Envy-freeness and relaxed stability: hardness and approximation algorithms},
  author={Krishnaa, Prem and Limaye, Girija and Nasre, Meghana and Nimbhorkar, Prajakta},
  journal={Journal of Combinatorial Optimization},
  volume={45},
  number={41},
  year={2023},
  publisher={Springer},
  url          = {https://doi.org/10.1007/s10878-022-00963-x}
}

@article{HIM16,
  author    = {Koki Hamada and
               Kazuo Iwama and
               Shuichi Miyazaki},
  title     = {The {H}ospitals/{R}esidents {P}roblem with {L}ower {Q}uotas},
  journal   = {Algorithmica},
  volume    = {74},
  number    = {1},
  pages     = {440--465},
  year      = {2016},
  url       = {https://doi.org/10.1007/s00453-014-9951-z}
}

@article{iwama201425,
  title={A 25/17-{A}pproximation {A}lgorithm for the {S}table {M}arriage {P}roblem with {O}ne-{S}ided {T}ies},
  author={Iwama, Kazuo and Miyazaki, Shuichi and Yanagisawa, Hiroki},
  journal={Algorithmica},
  volume={68},
  number={3},
  pages={758--775},
  year={2014},
  publisher={Springer},
url          = {https://doi.org/10.1007/s00453-012-9699-2}
}

@article{GS62,
 author = {D. Gale and L. S. Shapley},
 ISSN = {00029890, 19300972},
 url = {http://www.jstor.org/stable/2312726},
journal = {The American Mathematical Monthly},
 number = {1},
 pages = {9--15},
 publisher = {Mathematical Association of America},
 title = {College {A}dmissions and the {S}tability of {M}arriage},
 volume = {69},
 year = {1962}
}

@article{paluch2014faster,
  title={Faster and {S}impler {A}pproximation of {S}table {M}atchings},
  author={Paluch, Katarzyna},
  journal={Algorithms},
  volume={7},
  number={2},
  pages={189--202},
  year={2014},
  publisher={MDPI},
 url          = {https://doi.org/10.3390/a7020189}
}

@inproceedings{lam20191,
  title={A (1+ 1/e)-{A}pproximation {A}lgorithm for {M}aximum {S}table {M}atching with {O}ne-{S}ided {T}ies and {I}ncomplete {L}ists},
  author={Lam, Chi-Kit and Plaxton, C Gregory},
  booktitle={Proceedings of the Thirtieth Annual ACM-SIAM Symposium on Discrete Algorithms},
  pages={2823--2840},
  year={2019},
  organization={SIAM},
url          = {https://doi.org/10.1137/1.9781611975482.175}
}

@article{huang2015improved,
  title={Improved approximation algorithms for two variants of the stable marriage problem with ties},
  author={Huang, Chien-Chung and Kavitha, Telikepalli},
  journal={Mathematical Programming},
  volume={154},
  pages={353--380},
  year={2015},
  publisher={Springer},
url          = {https://doi.org/10.1007/s10107-015-0923-0}
}

@article{roth1984stability,
  title={Stability and {P}olarization of {I}nterests in {J}ob {M}atching},
  author={Roth, Alvin E},
  journal={Econometrica: Journal of the Econometric Society},
  pages={47--57},
  year={1984},
  url       = {https://doi.org/10.2307/1911460},
  publisher={JSTOR}
}

@article{Yokoi20,
  author    = {Yu Yokoi},
  title     = {Envy-{F}ree {M}atchings with {L}ower {Q}uotas},
  journal   = {Algorithmica},
  volume    = {82},
  number    = {2},
  pages     = {188--211},
  year      = {2020},
  url       = {https://doi.org/10.1007/s00453-018-0493-7},
  bibsource = {dblp computer science bibliography, https://dblp.org}
}

@article{Roth86,
 ISSN = {00129682, 14680262},
 URL = {http://www.jstor.org/stable/1913160},
 author = {Alvin E. Roth},
 journal = {Econometrica},
 number = {2},
 pages = {425--427},
 publisher = {[Wiley, Econometric Society]},
 title = {On the {A}llocation of {R}esidents to {R}ural {H}ospitals: {A} {G}eneral {P}roperty of {T}wo-{S}ided {M}atching {M}arkets},
 volume = {54},
 year = {1986}
}

@article{BFIM10,
  author    = {P{\'{e}}ter Bir{\'{o}} and
               Tam{\'{a}}s Fleiner and
               Robert W. Irving and
               David Manlove},
  title     = {The {C}ollege {A}dmissions problem with lower and common quotas},
  journal={Theoretical Computer Science},
  volume    = {411},
  number    = {34-36},
  pages     = {3136--3153},
  year      = {2010},
  url       = {https://doi.org/10.1016/j.tcs.2010.05.005},
 publisher={Elsevier}
}

@article{roth84,
  title={The {E}volution of the {L}abor {M}arket for {M}edical {I}nterns and {R}esidents: {A} {C}ase {S}tudy in {G}ame {T}heory},
  author={Roth, Alvin E},
  journal={Journal of political Economy},
  volume={92},
  number={6},
  pages={991--1016},
  year={1984},
  publisher={The University of Chicago Press},
url={https://doi.org/10.1086/261272}
}

@article{kiraly2011better,
	title={Better and {S}impler {A}pproximation {A}lgorithms for the {S}table {M}arriage {P}roblem},
	author={Kir{\'a}ly, Zolt{\'a}n},
	journal={Algorithmica},
	volume={60},
	number={1},
	pages={3--20},
	year={2011},
	publisher={Springer},
url          = {https://doi.org/10.1007/s00453-009-9371-7}
}

@article{Kiraly13,
  author    = {Zolt{\'{a}}n Kir{\'{a}}ly},
  title     = {Linear {T}ime {L}ocal {A}pproximation {A}lgorithm for {M}aximum {S}table {M}arriage},
  journal   = {Algorithms},
  volume    = {6},
  number    = {3},
  pages     = {471--484},
  year      = {2013},
  url       = {https://doi.org/10.3390/a6030471}
}

@inproceedings{swat/IrvingMS00,
  author       = {Robert W. Irving and David F. Manlove and  Sandy Scott},
  title        = {The {H}ospitals/{R}esidents {P}roblem with {T}ies},
  booktitle    = {7th Scandinavian Workshop on Algorithm Theory, {SWAT} 2000, Bergen, Norway},
  pages        = {259--271},
  publisher    = {Springer},
  year         = {2000},
  url          = {https://doi.org/10.1007/3-540-44985-X\_24}
}

@inproceedings{biro2010popular,
	title={{P}opular {M}atchings in the {M}arriage and {R}oommates {P}roblems},
	author={Bir{\'o}, P{\'e}ter and Irving, Robert W and Manlove, David F},
	booktitle={Algorithms and Complexity, 7th International Conference, {CIAC} 2010, Rome, Italy},
	pages={97--108},
	publisher    = {Springer},
        year         = {2010},
        url          = {https://doi.org/10.1007/978-3-642-13073-1\_10}
}

@article{NNRS2024popular,
  title={Popular critical matchings in the many-to-many setting},
  author={Nasre, Meghana and Nimbhorkar, Prajakta and Ranjan, Keshav and Sarkar, Ankita},
  journal={Theoretical Computer Science},
  volume={982},
  pages={114281},
  year={2024},
  publisher={Elsevier},
url          = {https://doi.org/10.1016/j.tcs.2023.114281}
}

@inproceedings{NNRS21,
  title={Popular {M}atchings in the {H}ospital-{R}esidents {P}roblem with {T}wo-{S}ided {L}ower {Q}uotas},
  author={Nasre, Meghana and Nimbhorkar, Prajakta and Ranjan, Keshav and Sarkar, Ankita},
  booktitle={41st {IARCS} Annual Conference on Foundations of Software Technology
                  and Theoretical Computer Science, {FSTTCS} 2021,
                  Virtual Conference},
  year={2021},
  url          = {https://doi.org/10.4230/LIPIcs.FSTTCS.2021.30}
}

@article{brandl2019two,
  title={Two {P}roblems in {M}ax-{S}ize {P}opular {M}atchings},
  author={Brandl, Florian and Kavitha, Telikepalli},
  journal={Algorithmica},
  volume={81},
  number={7},
  pages={2738--2764},
  year={2019},
  url= {https://doi.org/10.1007/s00453-019-00553-0},
  publisher={Springer}
}

@inproceedings{csaji2024AAMAS,
author = {Cs{\'a}ji, Gergely},
title = {A {S}imple 1.5-approximation {A}lgorithm for a {W}ide {R}ange of {M}aximum {S}ize {S}table {M}atching {P}roblems},
year = {2024},
url          = {https://dl.acm.org/doi/10.5555/3635637.3662890},
booktitle = {Proceedings of the 23rd International Conference on Autonomous Agents and Multiagent Systems, {AAMAS} 2024, Auckland, New Zealand},
pages = {409–415},
numpages = {7},
location = {Auckland, New Zealand}
}

@inproceedings{nasre2023critical,
  author       = {Meghana Nasre and
                  Prajakta Nimbhorkar and
                  Keshav Ranjan},
  title        = {Critical {R}elaxed {S}table {M}atchings with {T}wo-{S}ided {T}ies},
  booktitle    = {Graph-Theoretic Concepts in Computer Science - 49th International           Workshop, {WG} 2023, Fribourg, Switzerland},
  pages        = {447--461},
  publisher    = {Springer},
  year         = {2023},
  url          = {https://doi.org/10.1007/978-3-031-43380-1\_32}
}
\appendix

\newpage
\section{Error in the definition of uncertain proposal}
\label{append:kiralyM1}

We briefly describe Kir{\'a}ly's hospital-proposing algorithm for the many-to-one setting and give an example showing that requiring a hospital to be fully subscribed for its proposal to be uncertain can lead to an unstable matching. For details of the algorithm, we refer the reader to~\cite[Section 5]{Kiraly13}.

\subsection{Overview of Kir{\'a}ly's many-to-one algorithm}
The algorithm is proposal-based: hospitals make proposals, and residents accept or reject them. A hospital is said to be \emph{active} if it is under-subscribed and its preference list is non-empty.  An active hospital $h$ proposes to its \emph{favorite} resident $r$ on its preference list. Among residents at the same rank, $h$ gives priority to those who have not yet received a proposal from any hospital. If $r$ immediately rejects $h$, then $h$ deletes $r$ from its preference list $\prefh$.  

If a hospital's preference list becomes empty for the first time, it attains $*$ status and recovers its original list, retaining its current assignments. It resumes proposing from the beginning of the recovered list if it remains under-subscribed. A resident $r$ compares proposals according to its original ranking of the hospitals. If the hospitals are tied, $r$ prefers a starred proposal to an unstarred one.

A hospital $h$ is said to be \emph{uncertain} about a proposal to a resident $r$ if the following conditions hold: 
\begin{itemize}
    \item[(i)]  $h$ is fully subscribed,
    \item[(ii)] $h$ has not attained $*$ status, and
    \item[(iii)] there is another resident $r'$ still on $h$'s current list whom $h$ prefers to $r$ under its proposal rule. 
\end{itemize} 

A matched resident $r$ is \emph{precarious} if its current hospital is uncertain about its proposal to $r$. The above uncertainty conditions are evaluated with respect to the current matching. Thus, a hospital's uncertainty about an accepted proposal, and the corresponding resident's precariousness, may change during execution.

Upon receiving a proposal from a hospital $h$, the resident accepts it if it is unmatched, is precarious, or prefers the incoming proposal to the one it currently holds under the comparison rule above, accounting for starred status. If it accepts, it rejects its previous hospital, if any; otherwise, it rejects $h$.

If a hospital $h$ proposes to $r$, and later $r$ rejects $h$, then $h$ deletes $r$ from its preference list, unless $r$ was precarious immediately before rejecting $h$. If $h$ was uncertain about $r$ and $r$ rejects $h$, then $h$ keeps $r$ on its preference list. If the preference list of a hospital becomes empty for the second time, the hospital becomes inactive forever and never proposes again. This completes the description of the algorithm.

\subsection{Example illustrating the error}
The following example shows that the requirement that ``$h$ is fully subscribed'' in the definition of \emph{uncertain proposal} can lead to a matching that is unstable. 
Consider the following execution of the algorithm on the instance shown in Figure~\ref{fig:KiralyM1}:

\begin{itemize}
    \item$a_1$ proposes to $b_1$.  ($b_1$ accepts $a_1$. Here $a_1$ is not uncertain about this proposal, because $a_1$ is not fully subscribed).
    \item  $a_2$ proposes to $b_1$. ($b_1$ rejects $a_2$ because $b_1$ is not precarious, is fully subscribed, and strictly prefers its current match $a_1$ over $a_2$.)
    \item $a_3$ proposes to $b_4$. ($b_4$ accepts $a_3$.)
    \item $a_4$ proposes to $b_4$. ($b_4$ accepts $a_4$ and rejects $a_3$.)

    \item $a_1$ proposes to $b_2$. ($b_2$ accepts $a_1$. At this time, since (i) $a_1$ is fully subscribed and (ii) $b_3$ is tied with $b_1$ and $b_2$ and has not yet received any proposal, $a_1$ is uncertain about both $b_1$ and $b_2$. Hence, both $b_1$ and $b_2$ become precarious.)
    \item $a_2$ proposes to $b_5$. ($b_5$ accepts.)

    \item $a_3$ proposes to $b_1$. ($b_1$ rejects $a_1$ and accepts $a_3$ because $b_1$ was precarious. Now $a_1$ is no longer uncertain about its proposal to $b_2$, and hence $b_2$ is no longer precarious.)

    \item $a_1$ proposes to $b_3$. ($b_3$ accepts $a_1$. At this time all hospitals are fully subscribed, and the algorithm terminates.)
\end{itemize}

\begin{figure}
    \centering
  \begin{tabular}{|ll|ll|}
        \hline
        \multicolumn{2}{|l|}{\makecell{Preference list of hospitals}} & \multicolumn{2}{l|}{\makecell{Preference list of residents}} \\
        \hline
            $[2]\ a_1$ :& $(b_1, b_2, b_3, b_4)$      & $[1]\ b_1$: & $a_1, a_2, a_3$\\[2pt]
		$[1]\ a_2$ :& $b_1, b_5$ & $[1]\ b_2$: & $a_1$\\[2pt]
            $[1]\ a_3$ :& $b_4, b_1$ & $[1]\ b_3$:       &    $a_1$        \\[2pt]
            $[1]\ a_4$ :& $b_4$ & $[1]\ b_4$:       &    $a_4,a_3,a_1$        \\[2pt]
            & & $[1]\ b_5$:       &    $a_2$        \\[2pt]
        \hline
    \end{tabular}
    \caption{The quota of $a_1$ is 2, and all other vertices have a unit quota. The hospital $a_1$ has all its neighbors in a single tie, and the preferences of all other vertices are strict.}
    \label{fig:KiralyM1}
\end{figure}
The algorithm terminates with the matching 
$$
M = \{(a_1,b_2), (a_1,b_3), (a_2,b_5), (a_3,b_1), (a_4,b_4)\}.
$$
The pair $(a_2,b_1)$ blocks $M$.

In our definition of uncertain proposal, we do not require that the proposing vertex $a$ is fully subscribed. Under Definition~\ref{def:uncertainProp}, the first proposal from $a_1$ to $b_1$ is uncertain because $b_2$, for example, is unproposed, under-subscribed, and tied with $b_1$. Thus, $b_1$ would accept $a_2$ in the second step.

\section{Generalised Kir{\'a}ly's algorithm}\label{append:kiralyMM}

Algorithm~\ref{algo:23stableMM} gives the pseudocode for our adaptation of Kir{\'a}ly's algorithm~\cite{Kiraly13} to the many-to-many setting. Each vertex $a\in \AA$ proposes to its neighbors until $a$ is either fully subscribed or has proposed with the $*$ status to all vertices in its preference list. Throughout the algorithm, we use $|M(a)|$ to denote the number of vertices matched to $a$, irrespective of starred status. A vertex $a$ can propose $b$ at most three times -- at most twice without $*$ status and at most once with $*$ status. Line~\ref{alg1:upgrade} in Algorithm~\ref{algo:23stableMM} ensures that $a$ occupies at most one position at $b$ at any time.

Once a vertex $b\in\BB$ becomes fully subscribed, it remains so. If $b$ is already fully subscribed when it receives an unstarred proposal from $a$, then no neighbor tied with $b$ in $\prefa$ is both under-subscribed and unproposed by $a$; otherwise, the favorite-neighbor rule (Definition~\ref{def:favNbr}) would select such a neighbor before $b$. Hence, this proposal cannot be labeled uncertain. Since starred proposals are never labeled uncertain, we obtain the following observation.

\begin{observation}\label{obs:fullNotuncertain}
Once $b\in\BB$ becomes fully subscribed, no subsequent proposal to $b$ can be labeled uncertain.
\end{observation}

\begin{algorithm}
    \caption{A $\frac{2}{3}$-approximation to a maximum-size stable matching in $G=(\AA\cup\BB,E)$}
    \label{algo:23stableMM}
    \DontPrintSemicolon
    \SetAlgoLined
    Set $M=\emptyset$ and initialize a queue $Q$ with all $a\in\AA$\;
    Initially, every $a\in\AA$ is unstarred, and for each $a\in\AA$ all neighbors are unmarked\;

    \While{$Q$ is not empty}{
        let $a=\mathrm{dequeue}(Q)$\;
        \If{$a$ is unstarred}{
            \If{some neighbor in $\prefa$ is marked or unproposed by $a$}{
                let $b$ be the favorite neighbor of $a$, at rank $k$\;
                label uncertain proposals from $a$, if any, at rank $k-1$ as `not uncertain'\;
                let $\mathit{first}$ be true if $a$ has not yet proposed to $b$\;
                \textbf{if} $b$ is marked by $a$ \textbf{then} $a$ unmarks $b$\;
                $a$ proposes to $b$\;

                \If{$b$ is under-subscribed}{
                    $M=M\cup\{(a,b)\}$\;
                    \If{$\mathit{first}$ and there exists $b'\ne b$ at rank $k$ in $\prefa$
                        with $|M(b')|<q^+(b')$ that is unproposed by $a$}{
                        label $(a,b)$ as an uncertain proposal\;
                    }
                }
                \Else{
                    \If{$b$ holds an uncertain proposal from some $a'\in M(b)$}{
                        $M=(M\setminus \{(a',b)\}) \cup \{(a,b)\}$\;
                        $a'$ marks $b$\label{alg1:mark}\;
                        add $a'$ to $Q$ if $a'\notin Q$\;
                    }
                    \Else{
                        let $a'$ be a least-preferred partner in $M(b)$, accounting for starred status\;
                        \If{$a\succ_b a'$}{
                            $M=(M\setminus \{(a',b)\}) \cup \{(a,b)\}$\;
                            add $a'$ to $Q$ if $a'\notin Q$\;
                        }
                    }
                }
                \textbf{if} $|M(a)|<q^+(a)$ and $a\notin Q$
                \textbf{then} add $a$ to $Q$\label{alg1:underSubA11}\;
            }
            \Else{
                label all remaining uncertain proposals from $a$ as `not uncertain'\;
                $a$ attains $*$ status and is added to $Q$\tcp*{$a^*$ starts at the beginning of $\prefa$}
            }
        }
        \Else{
            \If{$a^*$ has not proposed to all vertices in $\prefa$}{
                let $b$ be the favorite neighbor of $a^*$\;
                $a^*$ proposes to $b$\;
                \If{$(a,b)\in M$}{
                    $M=(M\setminus\{(a,b)\})\cup\{(a^*,b)\}$\label{alg1:upgrade}\;
                }
                \Else{
                    let $a'$ be a least-preferred partner in $M(b)$, accounting for starred status\;
                    \If{$a=_b a'$ and $b$ holds an unstarred proposal from $a'$}{
                        $M=(M\setminus\{(a',b)\})\cup\{(a^*,b)\}$\;
                        add $a'$ to $Q$ if $a'\notin Q$\;
                    }
                }
                \textbf{if} $|M(a)|<q^+(a)$ and $a\notin Q$
                \textbf{then} add $a$ to $Q$\label{alg1:underSubA12}\;
            }
        }
    }
    \Return the underlying matching $M$, ignoring starred status\;
\end{algorithm}

\subsection{Correctness of the generalised  Kir{\'a}ly's algorithm}\label{sec:GenKiralyCorr}
To show the correctness, we prove that the matching $M$ output by Algorithm~\ref{algo:23stableMM} is stable and is a $\frac{2}{3}$-approximation to the maximum-size stable matching in $G$.

\begin{lemma}\label{lem:stableAlgo2}
The matching $M$ output by Algorithm~\ref{algo:23stableMM} is stable.
\end{lemma}
\begin{proof}
Assume for contradiction that $(a,b)$ is a blocking pair w.r.t. $M$. We consider three cases:
\begin{itemize}
    \item \textbf{$b$ is under-subscribed:} Since the number of partners of $b$ never decreases, $b$ was under-subscribed throughout the algorithm and accepted every unstarred proposal without rejecting any matched partner. A first rejection of a matched partner could not have been caused by a starred proposal either because the proposer was still matched to $b$ from its earlier unstarred proposal. Thus, the starred proposal would only have upgraded the existing pair.

    Since $(a,b)\notin M$, it follows that $a$ never proposed to $b$. Thus, $a$ is fully subscribed at termination; otherwise, it would have continued proposing. Moreover, since $b$ remained unproposed, the favorite-neighbor rule implies that $a$ never proposed to a neighbor ranked strictly below $b$. Hence, $a$ does not strictly prefer $b$ to any partner in $M(a)$, contradicting that $(a,b)$ blocks $M$.
    
    \item \textbf{$a$ is under-subscribed:} Since $a$ is under-subscribed at termination, it exhausted its preference list, including all marked neighbors, before attaining starred status. Consider its last unstarred proposal to $b$. Since $(a,b)\notin M$, $b$ rejected it immediately or rejected it later, possibly after an upgrade. This rejection was not due to uncertainty about $(a,b)$ because before promotion, such a rejection would force another unstarred proposal to $b$, and at promotion all remaining uncertainty labels are cleared. 
    
    To see why $b$ held no uncertain proposal at this rejection, consider the first starred proposal to $b$ from a non-partner, if any. Its proposer's last unstarred proposal to $b$ had already been rejected, and, by the argument above, not because it was uncertain. Since all earlier starred proposals to $b$ were upgrades, this earlier rejection involved an unstarred proposal, that is, $b$ either rejected the incoming proposal or replaced a least-preferred partner. In either case, $b$ was full and held no uncertain proposal, and it remained so by Observation~\ref{obs:fullNotuncertain}. 
    Thus, when $b$ rejected $a$, it was full and held no uncertain proposal. If it rejected $a$ immediately, it weakly preferred every partner to $a$. If it rejected $a$ later, then $a$ was a least-preferred partner and its replacement was at least as good as $a$ in $\prefb$. Later replacements could not worsen $b$'s least-preferred partner. Hence, $(a,b)$ does not block $M$.

    \item \textbf{Both $a$ and $b$ are fully subscribed:} Since $(a,b)$ blocks $M$, there is some $b'\in M(a)$ with $b\succ_a b'$. Thus, $a$ must have proposed to $b$ in its unstarred list before proposing to $b'$. Consider its last unstarred proposal to $b$. Since $(a,b)\notin M$, this proposal was rejected immediately or was rejected later, possibly after an upgrade. 
    
    The rejection could not have been due to uncertainty about $(a,b)$. Before going past the rank of $b$, such a rejection would have marked $b$ and forced $a$ to propose to $b$ again; after passing that rank, the proposal was no longer uncertain. As shown in the preceding case, $b$ therefore weakly preferred every partner it held after the rejection to $a$, and this remained true thereafter. Since $b$ is fully subscribed, $(a,b)$ cannot block $M$.

\end{itemize}
Thus, the matching $M$ output by Algorithm~\ref{algo:23stableMM} is stable.
\end{proof}

\begin{lemma}\label{lem:23stable}
Let $M$ be the matching returned by Algorithm~\ref{algo:23stableMM},
and let $M_{opt}$ be a maximum-size stable matching in $G$.
Then $|M|\ge \frac{2}{3}|M_{opt}|$.
\end{lemma}

\begin{proof}
    We prove this by showing no 1-length augmenting path\footnote{A 1-length augmenting path w.r.t. $M$ is an edge $(a,b)\notin M$ such that both $a$ and $b$ are under-subscribed in $M$.} or a 3-length augmenting path\footnote{A 3-length augmenting path w.r.t. $M$ is a 3-length path $\langle a',b,a,b'\rangle$ such that $(a,b)\in M$, $(a',b), (a,b')\notin M$ and both $a'$ and $b'$ are under-subscribed in $M$.} exists w.r.t. $M$ in $M_{opt}\oplus M$. Since the matching is stable, it is clear that there does not exist any 1-length augmenting path w.r.t. $M$ in $M_{opt}\oplus M$. Thus, we need to show that there is no 3-length augmenting path w.r.t. $M$ in $M_{opt}\oplus M$. 
    
    Suppose for contradiction that there is a 3-length augmenting path $\langle a_1, b, a, b_1\rangle$ such that $(a,b)\in M$ and the other two edges are in $M_{opt}$.    Clearly, $a_1$ remains under-subscribed in $M$ at the end of the algorithm. This implies that the proposal of $a_1^*$ was rejected by $b$. This further implies that $b$ is fully subscribed, and there is no uncertain proposal involving $b$ when $b$ rejected $a_1^*$. By Observation~\ref{obs:fullNotuncertain}, no proposal to $b$ is labelled uncertain afterwards. Also, note that $b_1$ is under-subscribed in $M$ and $(a,b_1)\notin M$. This implies that $a$ never proposed $b_1$, which further implies that $a$ did not exhaust proposing all vertices in $\prefa$ even without $*$ status. Therefore, $a$ did not achieve the $*$ status during the course of our algorithm. Since $b$ rejected $a_1^*$, we conclude that $a\succ_b a_1$, that is, $b$ ranks $a$ strictly better than $a_1$ in $\prefb$.

    Since $a$ did not propose $b_1$ during the course of Algorithm~\ref{algo:23stableMM}, $a$ is full even before proposing to $b_1$. This further implies that $b_1\not\succ_a b$, that is, $a$ ranks $b$ either strictly better than $b_1$ or at the same rank as $b_1$. Thus, $b \succeq_a b_1$. Now, we show that $b\succ_a b_1$. This immediately contradicts the stability of $M_{opt}$. 
    
    Suppose $b_1=_a b$ such that $b$ and $b_1$ are at $k^{th}$ rank in $\prefa$. Since $a$ is matched to a $k^{th}$-rank neighbor $b$ and an unproposed under-subscribed neighbor $b_1$ exists at the same rank $k$,  the proposal $(a,b)$ is labelled uncertain. Note that $a$ did not propose to any $(k+1)^{th}$-rank neighbor during the course of the algorithm because at least one $k^{th}$-rank neighbor $b_1$ remains unproposed by $a$ at the end of the algorithm. We consider the following two cases based on who among $a_1^*$ and $a$ proposed $b$ first.

\vspace{0.1in}

    \noindent\textbf{$a_1^*$ proposes to $b$ before $a$ proposes to $b$:} As the proposal from $a_1^*$ to $b$ is not the first proposal from $a_1$ to $b$, the proposal $(a_1^*,b)$ cannot be labelled uncertain. Three things can happen when $b$ receives the proposal from $a_1^*$ -- $(i)$ $b$ rejects $a_1^*$, $(ii)$ $b$ accepts $a_1^*$ and is fully subscribed and $(iii)$ $b$ accepts $a_1^*$ and is under-subscribed.  In the first two scenarios, $b$ cannot be the favorite neighbor of $a$ until $a$ has proposed to $b_1$. So, $a$ must propose $b_1$ before $a$ proposes $b$. The fact that $(a,b)\in M$, $a$ must have proposed to $b$, which implies $a$ also proposed to $b_1$. Since $b_1$ is under-subscribed, $b_1$ accepts $a$, and hence, $(a,b_1)\in M$ -- a contradiction. In $(iii)$, the proposal $(a,b)$ is labeled as an uncertain proposal because $b_1$ is under-subscribed and is unproposed by $a$. When $a$ proposes to $b$, $b$ must still be under-subscribed; otherwise, $a$ would choose the tied, under-subscribed, and unproposed neighbor $b_1$ instead. Thus, $b$ still holds $a_1^*$, since it cannot reject a partner before becoming fully subscribed. Since the proposal $(a_1^*,b)$ is not uncertain and $(a,b)$ is uncertain, $b$ must reject $a$ before rejecting $a_1^*$. The fact that $(a_1^*,b)\notin M$ implies that $b$ rejected $a_1^*$ and thus it also rejected $a$. Now, $a$ must propose $b_1$ before proposing to $b$. The fact that $(a,b)\in M$ implies that $a$ proposed to $b$ again, and hence it proposed to $b_1$.  Since $b_1$ is under-subscribed, $b_1$ accepts $a$, and $(a,b_1)\in M$ -- a contradiction.

\vspace{0.1in}

     \noindent\textbf{$a$ proposes to $b$ before $a_1^*$ proposes to $b$:} The proposal $(a,b)$ is an uncertain proposal because $b_1$ is under-subscribed and is unproposed by $a$.  The fact that $(a_1,b)\not\in M$ implies $b$ rejected $a_1^*$ ($a_1^*$ must have proposed to $b$ as it remained under-subscribed in $M$). But $(a_1^*,b)$ is not an uncertain proposal. Thus, $b$ must reject $a$ before rejecting $a_1^*$. The fact that $(a_1^*,b)\notin M$ implies that $b$ rejected $a_1^*$ and thus it also rejected $a$. Now, $a$ must propose $b_1$ before proposing to $b$. The fact that $(a,b)\in M$ implies that $a$ proposed to $b$ again, and hence it proposed to $b_1$.  Since $b_1$ is under-subscribed, $b_1$ accepts $a$, and $(a,b_1)\in M$ -- a contradiction.

     At each vertex, pair incident edges of $M\setminus M_{opt}$ with incident edges of $M_{opt}\setminus M$ until one of the two sets is exhausted. This decomposes the symmetric difference into alternating trails and cycles. Any trail with one more edge of $M_{opt}$ than of $M$ has both endpoints under-subscribed in $M$. By the argument above, such a trail has length at least five, so it contains $k$ edges of $M$ and $k+1$ edges of $M_{opt}$ for some $k\ge2$. Since $k+1\le 3k/2$, summing over all trails and cycles and adding back the common edges gives $|M_{opt}|\le \frac{3}{2}|M|$.
\end{proof}

\end{document}